\documentclass[aps,onecolumn,preprint,superscriptaddress,nofootinbib,floats]{revtex4}
\usepackage{amsmath,amssymb,color,mathrsfs, graphicx,verbatim,epsfig, bbm, wasysym}
\usepackage[hyperfootnotes=false]{hyperref}
\usepackage{slashed}
\allowdisplaybreaks

\def\lsim{\mathrel{\rlap{\lower4pt\hbox{\hskip1pt$\sim$}}
    \raise1pt\hbox{$<$}}}
\def\gsim{\mathrel{\rlap{\lower4pt\hbox{\hskip1pt$\sim$}}
    \raise1pt\hbox{$>$}}}

\newcommand{\gev}{{\rm GeV}}

\newcommand{\be}{\begin{eqnarray}}
\newcommand{\ee}{\end{eqnarray}}

\newcommand{\cO}{{\cal O}}
\newcommand{\met}{{\slashed E_T}}

\def\addresses#1#2{\hbox to \hsize{\@tablebox{#1}\hfil\@tablebox{#2}}}
\def\@tablebox#1{\vtop{\hsize=5in \begin{flushleft} #1 \end{flushleft}}}

\def\beq{\begin{equation}}
\def\eeq{\end{equation}}
\def\bit{\begin{itemize}}
\def\eit{\end{itemize}}
\def\beqa{\begin{eqnarray}}
\def\eeqa{\end{eqnarray}}

\def\MadGraph{{\tt MadGraph}}
\def\MadGraph5{{\tt MadGraph5}}

\def\stop{\tilde t}
\def\sbottom{\tilde b}
\def\gluino{\tilde g}
\def\mgluino{m_{\tilde g}}
\def\mstop{m_{\stop}}
\def\msbottom{m_{\sbottom}}

\begin{document}

\baselineskip 0.6cm

\begin{titlepage}

\thispagestyle{empty}

\begin{flushright}
%PAPER ID STUFF
\end{flushright}

\begin{center}

\vskip 2cm

{\Large \bf Boosting Searches for Natural SUSY with RPV \\ via Gluino Cascades}

\vskip 1.0cm
{\large  Zhenyu Han$^1$, Andrey Katz$^2$, Minho Son$^{3,4}$, and Brock Tweedie$^5$}
\vskip 0.4cm
{\it $^1$ Institute of Theoretical Science, University of Oregon, Eugene, OR 97403}\\ 
{\it $^2$ Center for the Fundamental Laws of Nature, Jefferson Physical Laboratory,\\ 
Harvard University, Cambridge, MA 02138} \\
{\it $^3$Department of Physics, Yale University, New Haven, CT 06511} \\
{\it $^4$Dipartimento di Fisica, Universit$\grave{a}$ di Roma ``La Sapienza" and}\\
{\it INFN Sezione di Roma, I-00185 Roma, Italy}\\
{\it $^5$ Physics Department, Boston University, Boston, MA 02215} \\
\vskip 1.2cm

\end{center}

\noindent  In the presence of even minuscule baryonic R-parity violation, the stop can be the lightest superpartner 
and evade LHC searches because it decays into two jets.  In order to cover this interesting possibility, we here 
consider new searches for RPV stops produced in gluino cascades.  While typical searches for gluinos decaying to 
stops rely on same-sign dileptons, the RPV cascades usually have fewer hard leptons, less excess missing energy, and 
more jets than R-parity conserving cascades.  If the gluino is a Dirac fermion, same-sign dilepton signals are also 
often highly depleted.  We therefore explore search strategies that use single-lepton channels, and combat backgrounds 
using $H_T$, jet counting, and more detailed multijet kinematics or jet substructure.  We demonstrate that the stop mass 
peaks can be fully reconstructed over a broad range of spectra, even given the very high jet multiplicities.  
This would not only serve as a ``double-discovery'' opportunity, but would also be a spectacular confirmation 
that the elusive top-partner has been hiding in multijets.

\end{titlepage}

\setcounter{page}{1}

\section{Introduction}
\label{sec:intro}

%%%%%%%%%%%%%%%%%%%%%%
%  Introduction
%%%%%%%%%%%%%%%%%%%%%%

We are currently at a very interesting crossroads for particle physics.  
The Higgs boson, or 
something very much like it, has recently been discovered at the 
LHC~\cite{Atlas:2012gk,CMS:2012gu}.  
At the same time, the LHC experiments have been engaging in a large 
variety of searches for the physics that stabilizes the Higgs potential, so far without any signs of new phenomena.  According to the usual logic, this is somewhat perplexing.  Cancellation of the top loop without fine-tuning requires the existence of a top-partner at the few-hundred GeV scale, and in most models this is accompanied by many other new colored particles.  The LHC is extremely efficient at producing colored particles, so if this scenario is correct, then the LHC is already creating top-partners and other new particles in abundance.  Where are they?

The non-observation of new physics has in particular made a major dent in the 
available parameter space for one of our best-motivated and well-defined 
stabilization mechanisms, supersymmetry (SUSY) (for summaries of experimental 
searches see~\cite{CMS:susysummary,Atlas:susysummary}).  
SUSY typically produces complicated cascade decay chains with copious 
production of jets, leptons, and---it is usually assumed---missing energy 
from an invisible lightest supersymmetric particle (LSP).  
One escape route for SUSY is to simply push up the masses of all superparticles 
beyond the LHC reach, though at the penalty of re-introducing the 
fine-tuning that it was invented to alleviate.  
A more attractive option from the perspective of naturalness
% and flavor physics
is what has come to be called ``natural SUSY'' or 
``effective SUSY''~\cite{Dimopoulos:1995mi,Cohen:1996vb} 
(see also~\cite{Barbieri:2010ar,Barbieri:2010pd}), where sleptons and 
first/second generation squarks are assumed to be very heavy, but the 
particles necessary to avoid Higgs sector fine-tuning remain relatively light.  
Because of its reduced particle content, natural SUSY can more easily evade 
constraints from generic SUSY searches~\cite{Kats:2011qh,Brust:2011tb,Papucci:2011wy}.
 But while, for example, relatively light stops and sbottoms can still be 
accommodated by the LHC data, targeted searches for these scenarios are gaining 
ground~\cite{Atlas:bMET,Atlas:2012si,Atlas:2012ar,Chatrchyan:2012wa,CMS:razorb}.

As our standard approaches to SUSY have thus far proven unrealized by nature, it 
is becoming clear that we must cast a wider net both in terms of our SUSY models and 
our strategies for searching for them.  This does not entail particularly baroque 
model-building, as even modest modifications to minimal SUSY models can radically change their phenomenology.  One of the simplest and well-known ways to do this is to turn on the baryon-number-violating and R-parity-violating (RPV) superpotential operator $W \sim u^cd^cd^c$.  If this operator has even a tiny nonvanishing coefficient, all supersymmetric decay chains are fated to end in jets instead of the usual missing energy.  This turns the LHC's virtue of copious colored particle production into a curse, as these signals can easily become lost in the noise of QCD multijet backgrounds.  Nonetheless, the LHC detectors are already proving themselves quite reliable for reconstructing detailed kinematic features using jets.  RPV signals are uniquely well-suited to this because they can contain multijet resonances, and more generally jet-lepton resonances.  Several papers have pointed out how some of these resonances might be 
uncovered~\cite{Desai:2010sq,Kilic:2011sr,Brust:2012uf,Evans:2012bf}, and searches are already underway~\cite{CMS:2012gw,ATLAS:2012dp}.  

R-parity violation is a well-motivated possibility from the perspective of 
natural SUSY~\cite{Brust:2011tb}.  Completely eliminating the sleptons and 
first/second generation squarks from the theory, we can view the remaining 
particle content as an effective field theory, and allow for a wide range of 
ignorance about physics above a scale beyond the LHC's direct reach, e.g. $\cO$(10~TeV).   
The theory's UV completion may not even be supersymmetric, as 
in~\cite{Sundrum:2009gv}.  The main argument for exact R-parity conservation, 
and the ensuing vanishing of all of the dimension-four RPV superpotential operators, 
is that it automatically ensures proton stability.  However, in a natural SUSY 
effective theory, there are no longer any guarantees that higher-dimensional 
R-parity-conserving operators are suppressed by a very high supersymmetric 
GUT scale, and we must impose additional symmetries such as exact baryon- or 
lepton-number conservation.  The requirement of R-parity conservation then becomes 
redundant, and we can consider turning on RPV couplings consistent with the more 
fundamental proton-stabilizing symmetries.  The size of these couplings is still 
constrained by flavor physics measurements (see~\cite{Barbier:2004ez} for 
review) and other indirect constraints that we briefly review below, but they are capable of mediating prompt decays at the LHC as long 
as they are larger than $\cO(10^{-7})$.

While various aspects of the impact of RPV on SUSY phenomenology at the LHC are 
being studied, perhaps the simplest scenario, highly motivated by naturalness yet 
very difficult to find at a hadron collider, is a stop LSP that decays directly 
to two jets.  Seemingly the most straightforward way to search for these particles is to pair 
produce them and look for pairs of dijet resonances within four-jet events.  
Searches of this type have been carried out by~\cite{CMS:coloron,Atlas:coloron} 
(and can be re-interpreted in terms of stop limits).  However, these searches are 
far too background-dominated to conclusively find or rule out stops.  Depending 
on the flavor structure of the $u^cd^cd^c$ superpotential operator, the decay 
may include a taggable $b$-jet that could be exploited in future four-jet searches, 
but this is far from guaranteed.

A search for $\stop \to jj$ in sbottom decays was proposed in~\cite{Brust:2012uf}, 
capitalizing on the presence of extra leptons from real or virtual $W$ emission 
in the $\tilde b \to \stop$ transition.  There it was shown that the stop may 
be visible as a dijet invariant mass peak rising above the backgrounds.  
Nonetheless, the coverage is spectrum-dependent, and many additional 
possible search options remain open for investigation.

In this paper, we turn our attention to searches for RPV stops produced in gluino 
decays.  If we take naturalness as a guide, then the gluino cannot be much heavier 
than a TeV, and should be within reach of the LHC~\cite{Brust:2011tb,Papucci:2011wy}.  
Producing stops in this way is potentially very useful for three reasons.  
First, gluino cross sections can be large and they tend to produce a lot of 
activity in their decays, making the stops easier to spot against background.  
Second, gluino decays most directly produce stops in association with top quarks.  
Since the stops lose their top quantum number in their RPV decay, becoming a 
fairly anonymous-looking dijet resonance, this may be one of the cleanest ways to 
establish that the resonance is in fact a top-partner.  Third, the gluino itself 
is an independent target in our searches for supersymmetry.  Of course, if this 
turns out to be the first search of this kind that yields a positive result, 
the impact of such a ``double-discovery'' would be quite dramatic.

Recently, there has been a study of gluino cascades into top quarks and RPV 
stops~\cite{Allanach:2012vj} that utilizes the same-sign dileptons that can 
result when a pair of gluinos decays into tops with equal charge.  Searches with 
same-sign dileptons are a classic way to look for gluinos, and exploit the 
fact that they are Majorana particles with no intrinsic sense of charge, as 
well as the fact that there are usually multiple chances for lepton production 
in their cascade decays.  Standard Model backgrounds to same-sign dilepton 
production are quite modest, allowing for a very clean search.  

Here, we will be pursuing a different strategy.  While same-sign dilepton searches can be expeditious for setting limits, it is not immediately clear that they are actually optimal when baryonic RPV decays are involved.  A standard R-parity conserving gluino decay via stop produces two top quarks (on- or off-shell), providing four opportunities for lepton production in a $\gluino\gluino$ pair event.  The overall branching ratio for same-sign dileptons is $\cO$(10\%).  The RPV cascade $\tilde g \to t \tilde t^* \to t (jj)$ instead provides one top quark from each side of the event, so the overall dilepton branching fraction is the same as in SM $t\bar t$, and only half of the dileptons are same-sign.  The same-sign dilepton branching fraction then shrinks to 2.5\%.  Some gain could be made by switching to an inclusive dilepton search strategy, combining same-sign and opposite-sign, with $t\bar t$+jets becoming the major limiting background for the latter.  However, as pointed out in~\cite{Lisanti:2011tm}, searches in the $l$+jets channel with high jet multiplicity also face $t\bar t$+jets as their leading background, but with much higher branching fractions (30\%).  Effectively, for the price of roughly doubling the relative background, we gain a factor of six in signal rate, and the latter is far more important when working near the edge of discovery or exclusion reach.  To capitalize on this, we therefore do not pursue dilepton-based searches, but focus on what might be possible with the far more plentiful $l$+jets.

There is also another reason to consider searches that do not rely on the presence of a low-background, same-sign dilepton component of the signal:  the gluino may be a Dirac fermion instead of a Majorana fermion.  This possibility, pointed out in~\cite{Hall:1990hq,Randall:1992cq,Fox:2002bu}, is well-motivated because it relaxes stringent constraints from low-energy tests such as FCNCs~\cite{Kribs:2007ac} and neutron-antineutron oscillation~\cite{Brust:2011tb}.  It also allows the gluino to be somewhat heavier (up to about $4m_{\tilde t}$) while still keeping the Higgs potential natural~\cite{Fox:2002bu}.  From the perspective of super-QCD, Dirac gluinos carry a well-defined fermion number, and can only be produced in gluino-antigluino pairs.  Their subsequent decays into tops and stops then depend sensitively on the exact spectrum.  In the limit where one species of purely-chiral stop dominates, gluinos will only decay into one specific sign of top quark, and antigluinos will decay into the other sign.  The subset of events where both tops decay leptonically will then be entirely {\it opposite}-sign.  More general spectra can lead to a more mixed composition of same-sign and opposite-sign, though an imbalance in favor of opposite-sign is generally preferred.  Our $l$+jets searches will be insensitive to such details, and therefore also uniquely suited to covering a broad range of possible spectra with Dirac gluinos.

Both Dirac and Majorana gluinos decaying to RPV stops will contribute an excess of $t\bar t$+jets, and can be revealed with high purity by placing stringent cuts on the total activity of these events.  The simplest version of such a search would demand a single hard, isolated lepton produced in association with a large number of jets and a large amount of scalar-summed jet $p_T$ ($H_T$), as was pursued in~\cite{Lisanti:2011tm}.  However, the signal also contains within it two dijet resonances.  These could be revealed if we can manage the combinatoric confusion presented by the additional jets from the top quark decays and from hard radiation.  Being able to explicitly reconstruct the $\stop \to jj$ peak is crucial to establish that we are producing a new particle in association with top quarks, a strong indication that it is indeed an RPV stop.  Having an additional highly-featured handle on the signal will also further improve $S/B$ and help reduce uncertainties associated with background modeling.  These are important benefits even from the perspective of just setting limits.

We will pursue two approaches to reconstructing the stop peak, which allow us to 
reliably combat combinatorics over a wide range of kinematics.  If the mass gap 
between the gluino and the stop is small, the top quarks receive little kinetic 
energy and the jets from the stop decays are usually amongst the hardest ones in 
the event.  We can then attempt a ``best pair of pairs'' approach similar to what 
is currently being used in searches for direct production of pairs of dijet 
resonances~\cite{CMS:coloron,Atlas:coloron}.  If the mass gap is instead large, 
then the tops and stops are both produced with fairly large boost.  
This reduces combinatorial uncertainties and renders the events suitable for 
analysis with jet substructure techniques 
(see~\cite{Abdesselam:2010pt,Altheimer:2012mn} for reviews).  
We will see that there 
is a sizable overlap between the regions of parameter space where either 
technique is effective, and that there are no ``weak spots'' in the sensitivity.
The combination of the two methods allows discovery-level sensitivity to roughly 1~TeV gluinos by the end of the 
2012 run, for almost any LSP stop mass.

Our paper is organized as follows. In the next section we will briefly review our assumption about RPV as well
as existing constraints on baryon-number violation, coming from flavor physics, other precision tests and 
cosmological considerations. We then go in detail through the relevant existing searches for SUSY and estimate the LHC constraints 
on our scenario.
In section~\ref{sec:strategies}, we describe our search strategies and estimate 
their discovery potential.
Finally, in section~\ref{sec:conclusions} 
we conclude.  An appendix contains details of our simulations.

\section{Constraints}
\label{sec:constraints}

%%%%%%%%%%%%%%%%%%%%%%
%  Constraints
%%%%%%%%%%%%%%%%%%%%%%

Throughout this paper we assume that lepton-number is perfectly conserved, 
thus preventing proton decay, 
while baryon-number is violated by the RPV superpotential operator
\beq
W = \lambda_{ijk} u^c_i d^c_j d^c_k~.  \label{eq:W}
\eeq
Note that $\lambda$ is antisymmetric in indices $j$ and $k$.  A stop LSP can 
then decay into two down-type jets through, e.g., $t^c s^c d^c$ or $t^c s^c b^c$.  
We will not make any assumptions about the flavor structure, and therefore do 
not attempt to capitalize on the possible presence of a $b$-jet in the decay.\footnote{However, we note that some theories can favor this, especially those motivated 
by minimal flavor violation~\cite{Csaki:2011ge,Berger:2012mm}.  Constraints 
on these scenarios might therefore be augmented by direct searches that 
have additional $b$-tagging demands, such 
as~\cite{Chatrchyan:2012wa,Chatrchyan:2012rg,Aad:2012pq}.}

Although proton decay is automatically prevented by lepton-number conservation, 
the presence of baryon-number violation induces several other effects that are 
highly constrained.  One of the simplest is $n - \bar n$ 
oscillation (see~\cite{Mohapatra:2009wp} for review). 
This would bound $\lambda \lesssim 10^{-5}$, but the constraint is easily avoided 
if the gluino mass is mostly Dirac.  There are also bounds from 
$K-\bar K$ mixing, which would require $\lambda \lesssim 10^{-2}$ if we assume 
mostly universal flavor structure.  Other indirect constraints from 
low-energy experiments tend to be 
weaker~\cite{Barbier:2004ez}.\footnote{Cosmologically, 
baryon-number violation can potentially wash out the baryon 
asymmetry produced in the early universe.  This is avoided if the coupling 
is less than $\cO(10^{-6})$.  However, one can also use (\ref{eq:W}) as a 
mechanism of electroweak-scale baryogenesis, requiring more sizable values 
of $\lambda$~\cite{Dimopoulos:1987rk,Cline:1990bw,Adhikari:1996mc}.}
 
At high-energy colliders, direct production of stops can be identified most 
simply by looking for pairs of dijet resonances.  The decays can be considered prompt as long as $\lambda\nolinebreak\gtrsim\nolinebreak10^{-7}$, a bound which can accommodate the above indirect constraints.  The LEP experiments searched for this process (which is quite analogous to $e^+e^- \to W^+W^- \to 4j$), placing a limit of $m_{\stop} > 90$~GeV for pure $\stop_R$~\cite{Heister:2002jc}, with a small allowed region at $m_{\stop} \simeq m_W$.  Similar searches have been done at the LHC~\cite{CMS:coloron,Atlas:coloron}, but there multijet production backgrounds swamp the signal.\footnote{No search of this type has been performed at the Tevatron, though it would likely be capable of making substantial gains.}

Hadron collider searches for gluinos decaying via stops usually make a set of 
implicit assumptions that lead to 
weakened sensitivity when the stops decay via RPV.  Nonetheless, these searches are still capable 
of setting limits, and in the remainder of this section we provide some estimates of what parameter space 
is still available for our own more targeted searches.

The usual assumption in gluino searches is that all cascades end in the emission of a neutralino LSP, 
giving an excess of missing energy beyond what could usually be provided by, for example, SM $t\bar t$.  
Some of the most powerful searches also capitalize on the fact that a gluino pair event can produce four top quarks, 
providing multiple opportunities for lepton emission and a high rate of multilepton events, 
including same-sign (SS) dileptons.  A canonical search of this type demands high 
missing energy, same-sign dileptons, and usually a handful of hard ($b$-)jets.  With this event 
selection, Standard Model backgrounds become very small, mainly populated by $t\bar t$ events where a 
lepton from $b$-quark decay is fairly hard and accidentally categorized as isolated, or an extra $W$ or $Z$ is emitted.  Our scenario does 
not produce four tops, but can still produce same-sign tops decaying to SS dileptons if the gluino is Majorana.

Generic SUSY searches have also been conducted using the opposite-sign (OS) dilepton channel, 
which is especially relevant for our Dirac gluino case.  In addition, both Dirac and Majorana gluinos 
might be picked up by the large variety of SUSY $l$+jets searches.  These are especially important for 
us to understand since our own proposed search strategy uses the $l$+jets channel.  Finally, as our 
signal is high-multiplicity and high-energy, we can also consider possible limits from searches 
for TeV-scale black holes and pairs fourth generation down-type fermions with $b'\bar b' \to (t W^-)(\bar t W^+) \to l\nu b\bar b 6j$.

\begin{figure}[tp]
\begin{center}
   \includegraphics[width=3.2in]{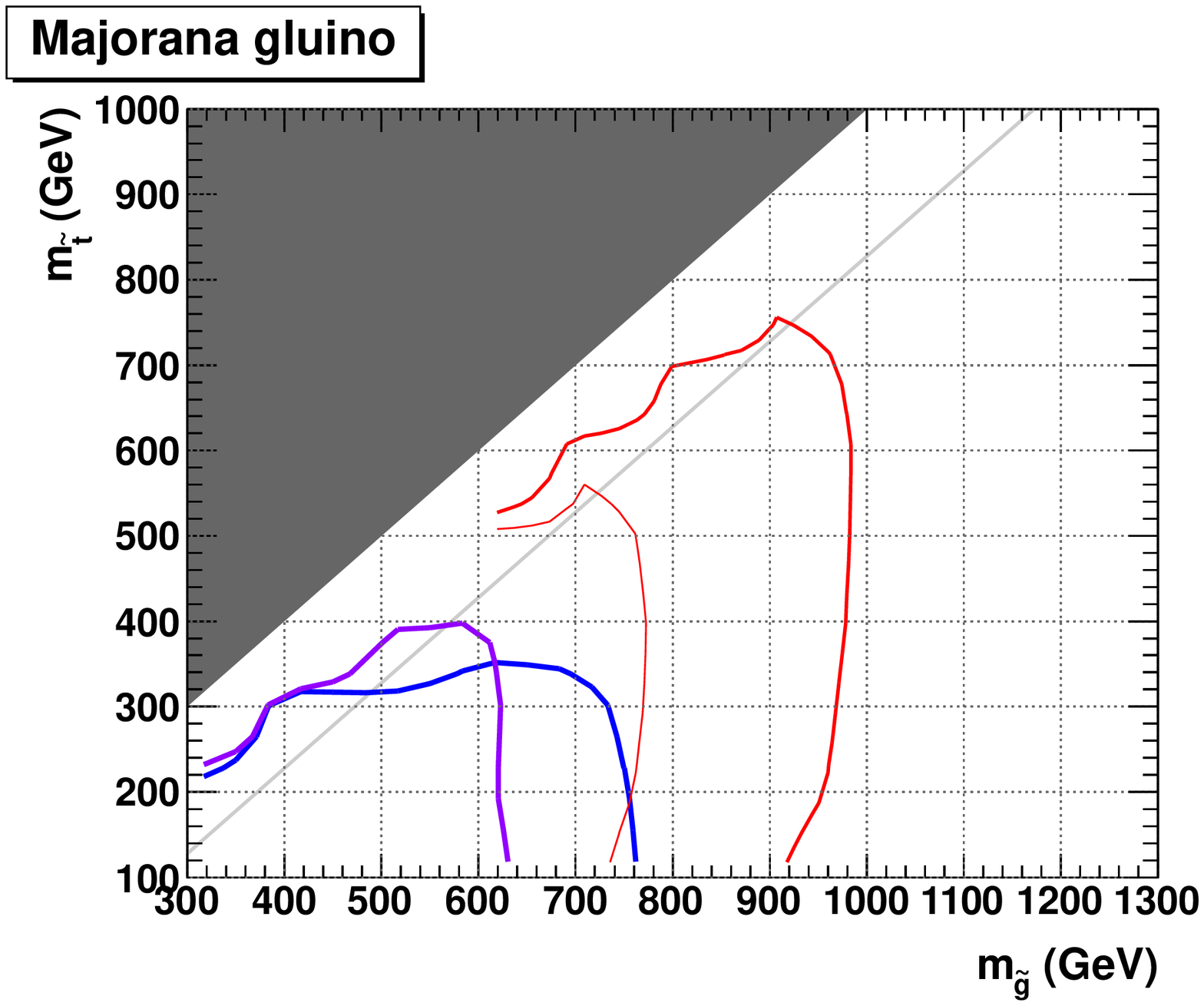}
   \includegraphics[width=3.2in]{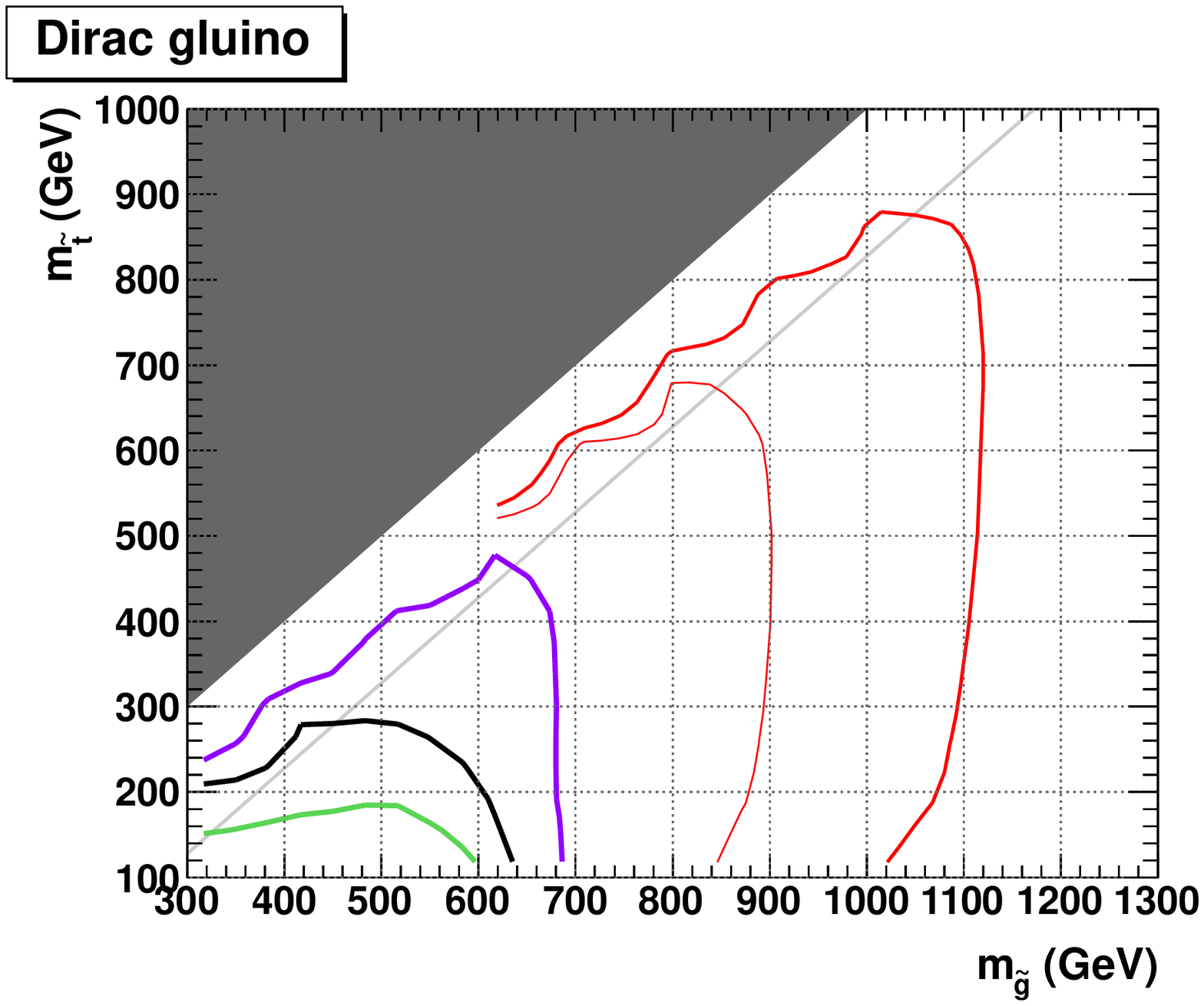}
\caption{Left: Constraints on a Majorana gluino.  The blue line is from the ATLAS same-sign dilepton search (LHC8, 6~fb$^{-1}$).  The purple line is from the ATLAS $b'$ search (LHC7, 1~fb$^{-1}$).  The red lines are our estimates for our $l$+jets $(N_j,H_T)$ style search (thin: assuming LHC7, 1~fb$^{-1}$;  thick: assuming LHC8, 5~fb$^{-1}$), with the boundary defined by $S/\sqrt{S+B} = 2$.   Right: Constraints on a Dirac gluino.  The green line is from the CMS opposite-sign dilepton SUSY search (LHC7, 5~fb$^{-1}$).  The black line is from the ATLAS black hole search (LHC7, 1~fb$^{-1}$).  The purple line is again ATLAS $b'$, and the red lines are again our $(N_j,H_T)$ counting estimates.  We do not consider regions with a $\gluino$ LSP, indicated with dark gray.  The light gray line indicates $m_{\gluino} = m_{\stop} + m_t$.}
\label{fig:constraints}
\end{center}
\end{figure}

Below, we describe some of the details of these searches.  We summarize our estimates of the most relevant limits in the $(m_{\gluino},m_{\stop})$ plane in Fig.~\ref{fig:constraints}, assuming $BR(\gluino \to t\stop) \equiv 1$.  These are supplemented by what could be obtained using our simplest high-multiplicity, high-$H_T$ search strategy described in section~\ref{sec:strategies}.  The conclusion is that Majorana (Dirac) gluinos at or above 760~GeV (690~GeV) are completely allowed by existing searches.  The strongest constraints occur when the stop is light, since then the top quark can carry more energy, and the lepton $p_T$ and $\met$ can be larger.  Majorana gluinos are more tightly constrained than Dirac gluinos when the stop is light because they always contribute to the SS dilepton channel.  Dirac gluinos could be much less constrained since they carry an approximately conserved type of fermion number, and in the simplest scenarios do not produce a SS dilepton signal.  However, they have two-times larger cross section, still produce an OS dilepton signal, and can also still be visible in black hole and $b'$ searches.  Limits from $l$+jets SUSY searches appear to be the weakest for both Dirac and Majorana.  In contrast to these non-dedicated searches, our proposed $l$+jets search should already be capable of ruling out 1~TeV gluinos with 5~fb$^{-1}$ of data at LHC8.

%----------------------------------
\subsection{Same-sign dileptons}
%----------------------------------

At present, the most powerful search in this category uses about 6~fb$^{-1}$ of data from LHC8, and has been performed by ATLAS~\cite{Atlas:ssdilep}.  This search demands two SS leptons produced in association with at least four jets of $p_T > 50$~GeV and 150~GeV of $\met$.  In a simplified R-parity conserving (RPC) model with a gluino decaying via off-shell stop into two tops and a light LSP neutralino, gluinos of mass less than 920~GeV are ruled out.  

Our gluino pair production signal produces two top quarks in association with four hard jets from the RPV stop decays.  It can contribute to the SS dilepton signal if both tops have the same charge and both decay semileptonically.  This is guaranteed to occur if the gluino is Majorana, since it then has no sense of fermion number, and can decay with equal probability into $t\tilde t^*$ and $\bar t \tilde t$.  The branching fraction of a complete event into SS dileptons is nonetheless quite small, as we pay the usual penalty of 5\% for dileptonic tops, times an extra factor of 1/2 to account for the probability that the tops have the same charge.  Subsequently, the requirement of four hard jets is almost trivially satisfied.  The $\met$ requirement is most easily met if the stop is somewhat light relative to the gluino, so that the tops can pick up a large fraction of the available energy in the decay and boost up the leptonically-decaying $W$-bosons.  

The upper limit on a new physics signal is 6.3 events.  For a Majorana gluino with $m_{\gluino} = 600$~GeV and $m_{\stop} = (100,200,400)$~GeV, we estimate that ATLAS would have observed (25,22,3) extra events.\footnote{We have validated our own Monte Carlo with respect to ATLAS by studying predicted kinematic distributions and event counts for their simplified RPC model.}  The smallest allowed $m_{\stop}$ is roughly 350~GeV for this gluino mass.  Note that the limits become weaker for heavier stops, since then the energy apportioned to the top quark is squeezed out.

The same channel for the same RPV model was also studied in~\cite{Allanach:2012vj}, re-interpreting SS dilepton searches from LHC7 with up to 2~fb$^{-1}$ of data.  They found gluino mass limits of 550~GeV, roughly independent of the stop mass, in the case where the RPV coupling is unity and mainly in regions of parameter space where the top or stop is off-shell.\footnote{We point out one subtlety here.  The behavior in the region $m_{\stop} < m_{\gluino} < m_{\stop} + m_t$ may be highly dependent on the RPV coupling strength.  For very small RPV coupling (e.g., $\lesssim 10^{-5}$), such as we are implicitly assuming to avoid precision constraints, the top will almost always be off-shell, and the stop on-shell.  The energy available to the $l\nu b$ system will therefore completely shut off as $m_{\gluino} \to m_{\stop}$ from above, eliminating the leptonic part of the signal and making limit-setting impossible.  However, for $\cO(1)$ RPV coupling, such as it was assumed in~\cite{Allanach:2012vj}, the stop can also go off-shell with appreciable rate, allowing the lepton to still carry some energy and nontrivial limits to be set.  When $m_{\gluino} < m_{\stop}$, this subtlety is sidestepped as the stop is then {\it forced} off shell, and the lepton can be energetic regardless of the coupling.  However, since we are assuming a stop LSP for naturalness reasons, this case formally falls outside the scope of our paper.}  The fully on-shell region which we have checked, with a more recent analysis from ATLAS, extend these results.

%--------------------------------------
\subsection{Opposite-sign dileptons}
%--------------------------------------

CMS has an OS dilepton search for generic RPC SUSY with 5~fb$^{-1}$ of LHC7 data~\cite{Chatrchyan:2012te}.  This analysis further requires two hard jets, and places a small variety of $H_T$ and $\met$ cuts defining four signal regions.  ($H_T$ is defined by summing over jets.)  For our purposes, the ``high $H_T$'' region is the most constraining, as our signal easily produces large amounts of $H_T$, but not necessarily large amounts of $\met$.  The cuts are $H_T > 600$~GeV and $\met > 200$~GeV.

This analysis places some constraints on Majorana gluinos with RPV stops, but they are much weaker than those from SS dilepton searches.  For Majorana gluinos, the rates in OS and SS are identical, but the backgrounds in the latter are significantly smaller.  Dirac gluinos, on the other hand, might only contribute to OS.  Assuming its decay is dominated by a single species of mostly-LH or mostly-RH stop, an approximate fermion number can be defined that is conserved throughout both the production and decay.  The sensitivity is also much better for Dirac than Majorana, for two reasons.  First, Dirac production cross sections are twice as big.  Second, we assume that dileptonic Dirac gluino pairs are always OS, whereas for Majorana they are only OS half of the time.  The total OS cross section for Dirac is therefore four times larger than for the equivalent Majorana model.

CMS places an upper limit of 23 new physics events in its high $H_T$ signal region.  For a Dirac gluino with $m_{\gluino} = 600$~GeV and $m_{\stop} = (100,200,400)$~GeV, we estimate that CMS would have observed (24,18,1.2) extra events.  The smallest allowed $m_{\stop}$ is therefore close to 100~GeV for this gluino mass.

%---------------------------
\subsection{Single lepton}
%---------------------------

Searches for SUSY in single-lepton channels can be nontrivial to interpret for our models.  Our signal 
essentially looks like $t\bar t$+jets, which occurs plentifully in the SM and serves as one of the 
dominant SUSY backgrounds.  SUSY searches therefore tend to craft cuts that either efficiently eliminate 
$t\bar t$+jets from the outset or attempt to fit it away using control regions.  Most of the ATLAS searches 
fall into the former category, by demanding that $m_T(l,\met)\nolinebreak > \nolinebreak 100$~GeV.  
Since this severely degrades our signal along with the SM $t\bar t$+jets, we do not study these searches 
in detail.  However, CMS has four recent searches which do not place such a cut.

Of all of these, probably the most robust for our purposes is~\cite{CMS:lmettemp}.  The strategy is to normalize the backgrounds in a low-$H_T$, low-$\met$ control region, extrapolate to high-$H_T$ and high-$\met$ using Monte Carlo, and count.  For this analysis, the control region should indeed be more highly purified in backgrounds relative to the signal regions, at least for $m_{\gluino} \gsim 400$~GeV.  The final analysis cuts demand at least three hard jets, $H_T > 750$ or 1000~GeV, $\met > 250$, 350, or 450~GeV, and various numbers of $b$-tags.  We find the best sensitivity in the signal regions with either of the two $H_T$ cuts, the smallest $\met$ cut, and no requirements on $b$-tags.  The strongest constraints are obtained for Dirac gluinos, due to their higher cross section.  However, the search proves to be strictly weaker than the OS dilepton.

The other three searches are likely incapable of placing stronger limits.  One search~\cite{CMS:ANN} uses an artificial neural network trained on the RPC model LM0, for which $m_T(l,\met)$ is actually the most powerful individual discriminating variable.  Another~\cite{CMS:lmet} looks for deviations in the $\met/\sqrt{H_T}$ spectrum, again normalizing backgrounds using control regions.  However, these include a high-$H_T$ and low-$\met/\sqrt{H_T}$ control region, which would actually be more signal-pure than their high-$H_T$ and high-$\met/\sqrt{H_T}$ signal region.  Even ignoring this issue completely, the final sensitivity would still be weaker than~\cite{Chatrchyan:2012te}.  Finally, the most recent search~\cite{CMS:PAS-SUS-12-010} attempts to exploit differences in the relationship between the lepton and $\met$ in SM backgrounds versus RPC SUSY signals.  For example, in $t\bar t$+jets, harder $\met$ tends to be associated with harder leptons.  Exactly the same relationships occur for our signal.  But again completely ignoring this issue, and just checking the counts in the signal regions, we expect sensitivity at most comparable to the OS dilepton search above.

%-----------------------------------
\subsection{Black hole searches}
%-----------------------------------

If TeV-scale black holes exist, they should be produced at the LHC and decay into high-multiplicity final states of jets, leptons, photons, and neutrinos.  Two of the most recent such searches are~\cite{CMS:BH} and~\cite{Atlas:BH}.  Neither of these searches places stringent requirements on $\met$, and only the second requires the presence of a lepton.  They can therefore be fairly efficient at capturing the gluino signal, even in the all-hadronic mode in the former case. 

In~\cite{CMS:BH}, a purely data-driven search by CMS using 3.7~fb$^{-1}$ at LHC8, the only discriminating variables are the inclusive number $N$ of reconstructed objects (jets, leptons, and photons) with $p_T > 50$~GeV, and the $S_T$ formed by summing all of these $p_T$'s (also adding $\met$ if it exceeds 50~GeV).  They consider events with $N \ge 3$--8 and $S_T > 1800$~GeV.  The $S_T$ shape is assumed to be independent of $N$, and is fit using $N = 2$, which is mainly dijets.  The predictions for $d\sigma/d\,S_T$ for the different $N$ cuts are then independently normalized using the data observed in the range $S_T = [1800,2200]$~GeV.  Model-independent results are presented for the total allowed $\sigma\times$(acceptance) of any new physics contribution as a function of $N$ and $S_T$ cuts.  The strongest bounds on our signal would come from the $N \ge 8$ region.  For example, an 800~GeV Dirac gluino with 300~GeV stop would contribute about 30 events, or 8~fb, for $S_T > 2200$~GeV.  The upper limit is about 6~fb, suggesting that gluinos up to 800~GeV may already be ruled out.  However, we point out an important caveat to this interpretation.  Like the SM, the gluino signal is a steeply-falling function in this region of high-$S_T$, and would contribute to the shape template normalization.  In our example model, the signal contribution in the CMS control region would be 70 events, whereas the number observed is roughly 150.  
We therefore do not feel that the claimed cross section limits can be robustly applied to this gluino model, as the signal highly contaminates the control region.  Heavier gluinos could be easier to discriminate in shape, but contribute much smaller rates.  A more careful $S_T$ shape analysis in high-multiplicity events could be useful here.

In~\cite{Atlas:BH}, a search by ATLAS using 1~fb$^{-1}$ of LHC7 data, the signal region requirements are at least one lepton with $p_T > 100$~GeV, at least two other jets or leptons with $p_T > 100$~GeV, and lower bounds on the scalar-summed $p_T$ over all jets and leptons varying from 700~GeV to 1500~GeV.  Backgrounds are normalized in low-summed-$p_T$ control regions.  The modest requirements on the number of reconstructed objects makes this search less sensitive than it otherwise might be for gluinos, but it can still place a nontrivial limit.  For 600~GeV gluinos with $m_{\stop} < 300$~GeV, we predict more than 30 events with summed-$p_T > 1200$~GeV, whereas ATLAS places an upper limit of 33 events.  This search is therefore stronger than CMS's 5~fb$^{-1}$ OS dilepton search.

%-----------------------------------------------------------------------
\subsection{ATLAS $b'$ search}
%-----------------------------------------------------------------------

Interestingly, ATLAS's fourth generation down-type fermion search~\cite{ATLAS:2012aw} puts one of the most stringent existing constraints on 
Dirac gluinos.\footnote{We are grateful to Tobias Golling for pointing out the relevance of this search for our signal.}
This search looks for pair-produced $b'$ fermions, each of which decays into a top and $W$.  Four $W$'s are ultimately produced
from the two $b' \to Wt$ and $t\to Wb$ decays.  The search considers the channel where exactly one of these $W$'s produces 
an observable, isolated lepton, and the others decay hadronically (or into taus or otherwise unobserved leptons).  Since the missing energy in signal events is coming only from a neutrino, this analysis 
imposes very mild cuts on $\met$ and $m_T$.

The nominal signal is one lepton plus eight jets, six of which can in principle be clustered into on-shell $W$'s.  In practice, the hadronic $W$'s, which are somewhat boosted in the heavy $b'$ decay, are identified by looking for pairs of jets with $\Delta R(j,j) < 1.0$ and $m(jj) = [70,100]$~GeV.  The events are binned according to the number of jets and the number of identified hadronic $W$'s.  We find that our gluino signal is most efficiently captured by the tightest bin, demanding at least eight jets and at least two $W$'s reconstructed out of these jets.  For the bulk of the parameter space constrained by this search, the acceptance is $\cO(20\%)$.  (For $m_{\stop} \simeq 100$~GeV, the search can even pick up the stops as ``$W$'s''.)

We show the exclusion of this search in Fig.~\ref{fig:constraints} (along with all of the most relevant searches discussed above). The resulting constraints are strong, 
excluding Dirac guinos up to almost 690 GeV for a light stop.  We note that the search has 
only been carried out up to a luminosity of 1~fb$^{-1}$ at 7~TeV.  However, any improvement of these bounds will probably
be very moderate as more data is added. This is because the  error is already largely 
dominated by systematics, mainly from the theoretical uncertainty on $t \bar t+$jets production.

The $b'$ search shares some similarities with the dedicated $\gluino \to t\stop$ searches that we propose in the next section.  Some differences are that we will also take advantage of the high-$H_T$ of gluino events and actively search for the stop mass peaks.  We will also replace the simple hadronic ``$W$-tagging'' with more sophisticated hadronic ``top-tagging.''  As can already be seen in Fig.~\ref{fig:constraints}, even our simplest dedicated search can in principle make major gains.

\section{Search Strategies and Results}
\label{sec:strategies}

%%%%%%%%%%%%%%%%%%%%%%
%  Reconstruction techniques
%%%%%%%%%%%%%%%%%%%%%%

While each LSP stop from a gluino pair always decays to two jets, the associated 
top can either decay semileptonically or hadronically.  We will focus on events
where one top decays semileptonically, providing us a clean trigger and suppressing multijet backgrounds, while maintaining a high branching fraction.
A simple search strategy is then to perform a survey over jet multiplicity, $N_j$, and event $H_T$, where
we define the latter as
\beq
H_T \,\equiv\, p_T(l) + \sum_i p_T(j_i)~.  \label{eq:HT}
\eeq
The main backgrounds are $t\bar t$+jets and $W$+jets.  These tend 
to occupy low ($N_j$,$H_T$), whereas the signal occupies high ($N_j$,$H_T$):
$N_j$ tends to be larger than six, and $H_T$ is strongly correlated with $2m_{\gluino}$.  Therefore,
our first approach will be to simply choose $N_j$ and $H_T$ cuts to maximize
the significance, very similar to what was proposed in~\cite{Lisanti:2011tm}.  
As we will show, gluinos up to roughly 1~TeV are discoverable
at the 8~TeV LHC using this simple method, neglecting possible systematic errors.

We will then show that it is possible to reconstruct the $\stop \to jj$ peaks in the signal, despite the nontrivial combinatorics in a high-$N_j$ environment.  The appropriate strategy depends on the mass hierarchy between the gluino and the stop.  When the masses are comparable, the stop is fairly heavy and the top quark receives fairly little kinetic energy in the gluino decay.  The jets from the stop decay then tend to be well-separated and higher in energy than the jets from top quark decay (or additional radiation jets).  We will apply a traditional jet reconstruction strategy to this case, and break the combinatoric ambiguities by identifying the two dijet subsystems that are closest in mass to each other, choosing only amongst the hardest handful of jets in the event.  When the stop mass is instead much smaller than the gluino mass, possibly even smaller than the top quark mass, both the stop and the top quark will be produced with appreciable boost.  A strategy based on jet substructure then becomes appropriate to identify ``stop-jets'' and distinguish these from top-jets.  Picking the best of these two procedures, and supplementing our $(N_j,H_T)$ cuts with a mass window cut centered on the stop mass, we obtain similar naive discovery potential as with $(N_j,H_T)$ alone.  However, this comes with better shape discrimination, enhanced $S/B$, and of course a much clearer picture of what is being produced.

Below, we will mainly work in a simplified model where $BR(\gluino \to t \stop) \equiv 1$.
The details of our Monte Carlo simulations and event reconstruction can be found in the appendix.  Note that we do not utilize any detector simulation, since a large amount of irreducible physics noise is already induced by combinatoric uncertainties and showering effects.  (Introducing a detector model does not change our results.)
We also find that our results are fairly insensitive to whether we are working with a Majorana or Dirac gluino, up to the fact that the latter's cross section is 2 times larger than the former.

%--------------------------------------
\subsection{$(N_j,H_T)$ cut-and-count analysis}
\label{subsec:cutandcount}
%--------------------------------------

\begin{figure}[t]
\begin{center}
\epsfxsize=0.44\textwidth\epsfbox{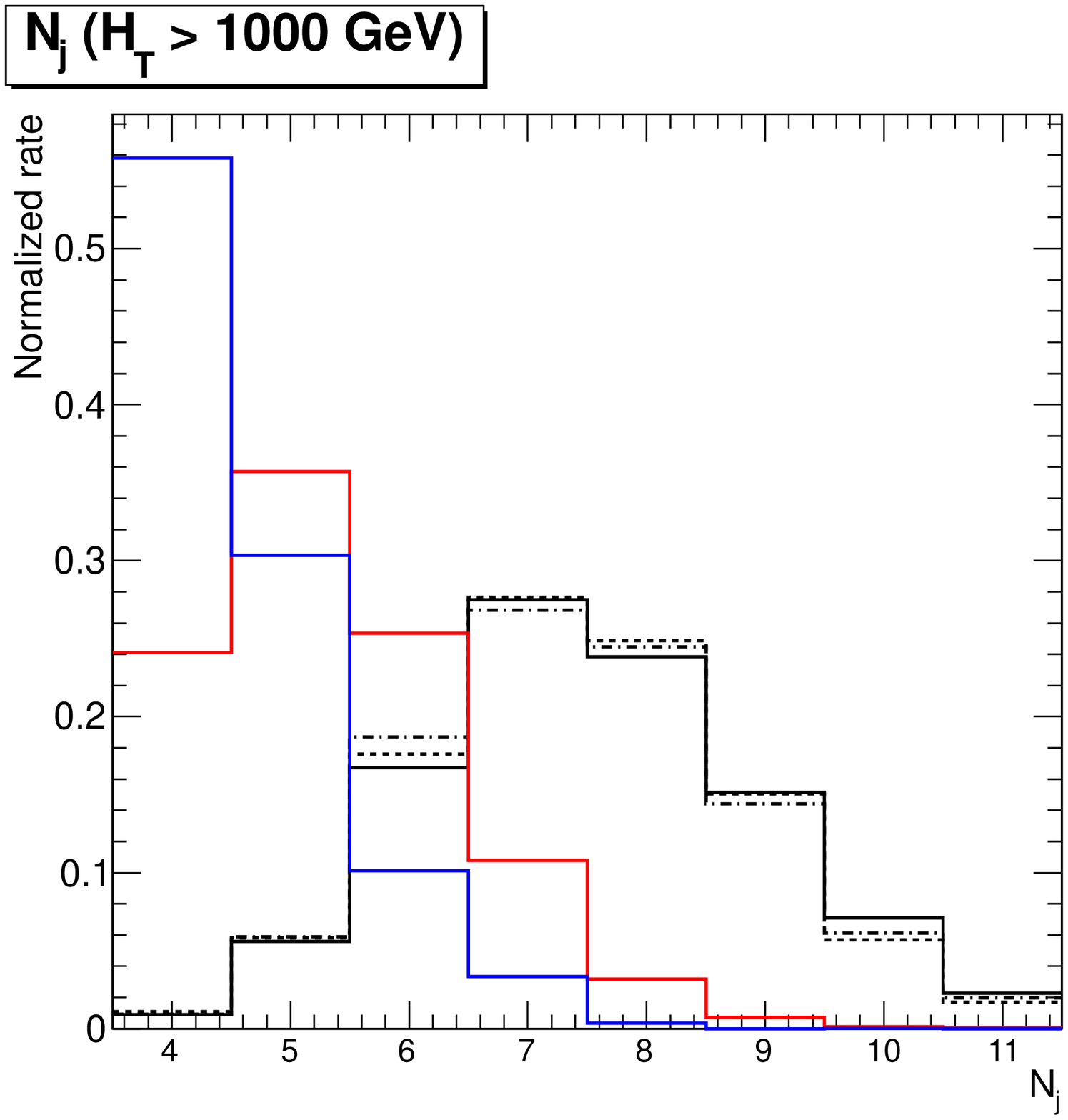}
\epsfxsize=0.44\textwidth\epsfbox{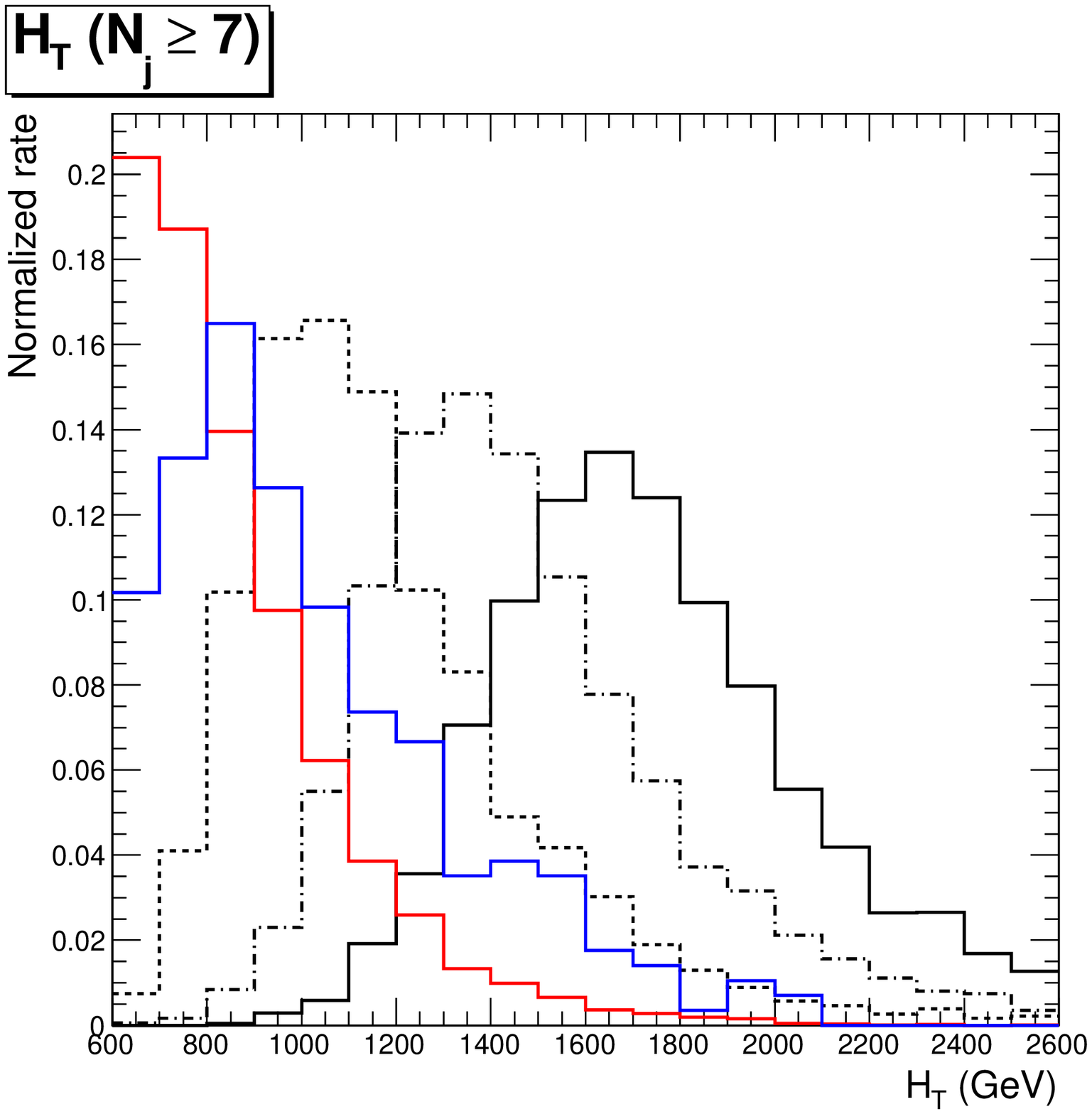}
\caption{Left: $N_j$ distributions of signals for $m_{\tilde g} =$ 600~GeV (black dotted), 
800~GeV (black dashed), and 1000~GeV (black solid) with $m_{\tilde t} = 1/2\times m_{\tilde g}$, 
and backgrounds $t\bar t$+jets (solid red) and $W$+jets (solid blue). 
Right: $H_T$ distributions of the corresponding signals and backgrounds.  The gluinos used for these plots are Dirac, but we obtain equivalent distributions with Majorana.}
\label{fig:HTandNj}
\end{center}
\end{figure}

We show the jet multiplicity distribution of high-$H_T$ events for 
several different gluino masses and for SM backgrounds in 
the left panel of Fig.~\ref{fig:HTandNj}. The SM background samples consist of $t\bar t$+jets matched up to two jets, and $W$+jets matched up to four jets, as described in the appendix.  As expected, the signal events 
occupy a higher $N_j$ region than the backgrounds. We also see that the 
$N_j$ distribution is insensitive to $m_{\tilde g}$, and it turns out 
that the cut $N_j \ge 7$ allows us to optimize the search significance over the entire range of 
gluino masses and stop masses that we consider.  Therefore, we first fix the cut on jet multiplicity 
to be $N_j \ge 7$.  After fixing this cut, we show the $H_T$ distributions 
in the right panel of Fig.~\ref{fig:HTandNj}, where we 
see that a cut on $H_T$ will further increase the signal/background ratio. 
For simplicity, we fix the $H_T$ cut such that $\sim$80\% of the signal events 
 are kept (after the $N_j\ge 7$ cut).  One can further optimize the set of 
$(N_j,H_T)$ cuts, but we will not pursue it in this work.

%------------------------------------------------
\subsection{Bump hunting with traditional jets}
\label{subsec:tradjetanalysis}
%------------------------------------------------

Our traditional jet analysis is appropriate when the jets from the stop decay are well-separated, and are the dominant jets in the event.  Of the seven or more jets in the event, the chances of randomly picking two jets from the same stop decay are small.  We therefore exploit the fact that there are two stops in the event, by searching for two pairs of jets with similar invariant masses and taking their average. Similar techniques were introduced in~\cite{Kilic:2008ub} and implemented in LHC searches for pair-produced dijet resonances~\cite{CMS:coloron,Atlas:coloron}. 

\begin{figure}[t]
\begin{center}
\epsfxsize=0.27\textwidth\epsfbox{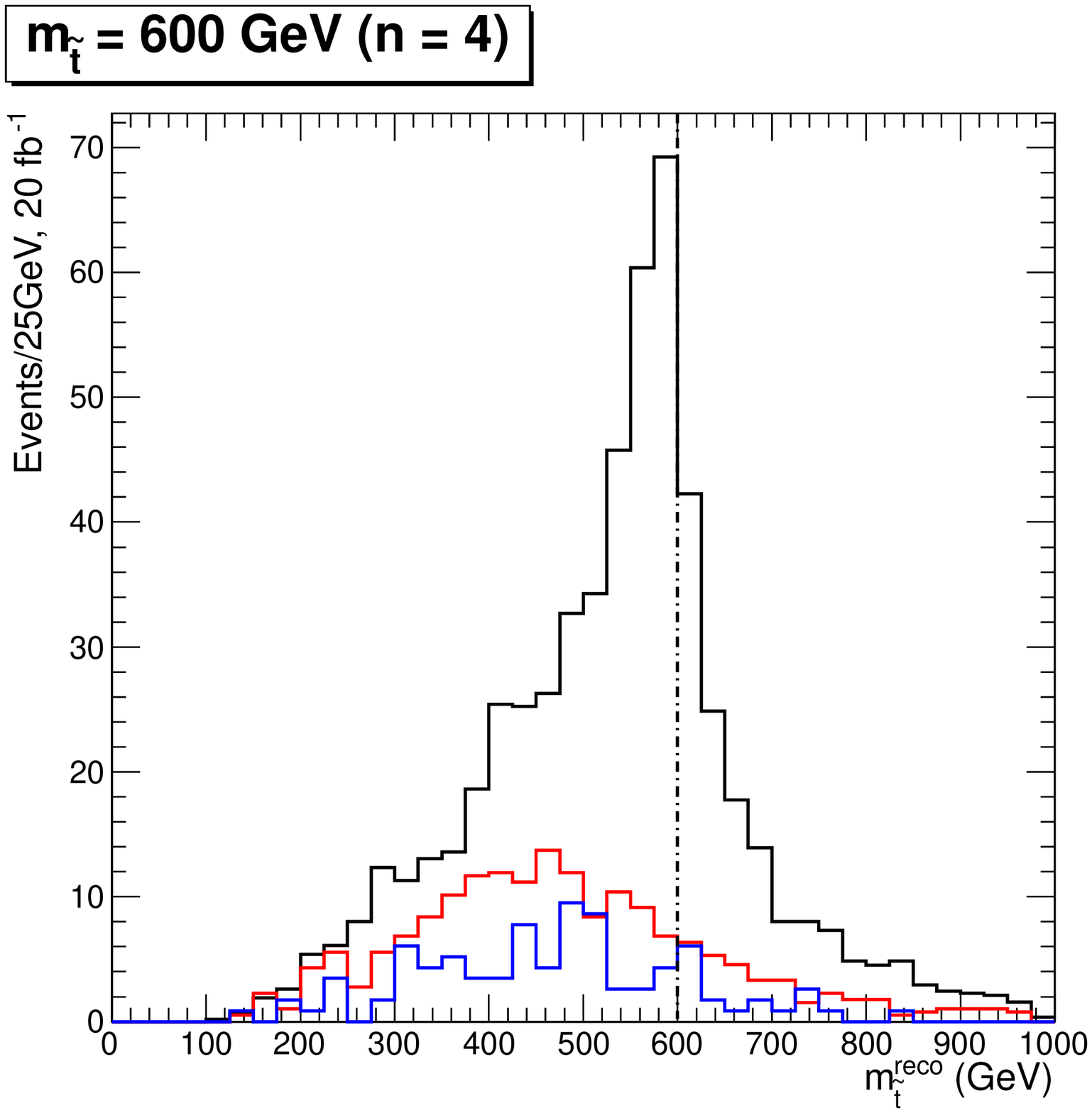}
\epsfxsize=0.27\textwidth\epsfbox{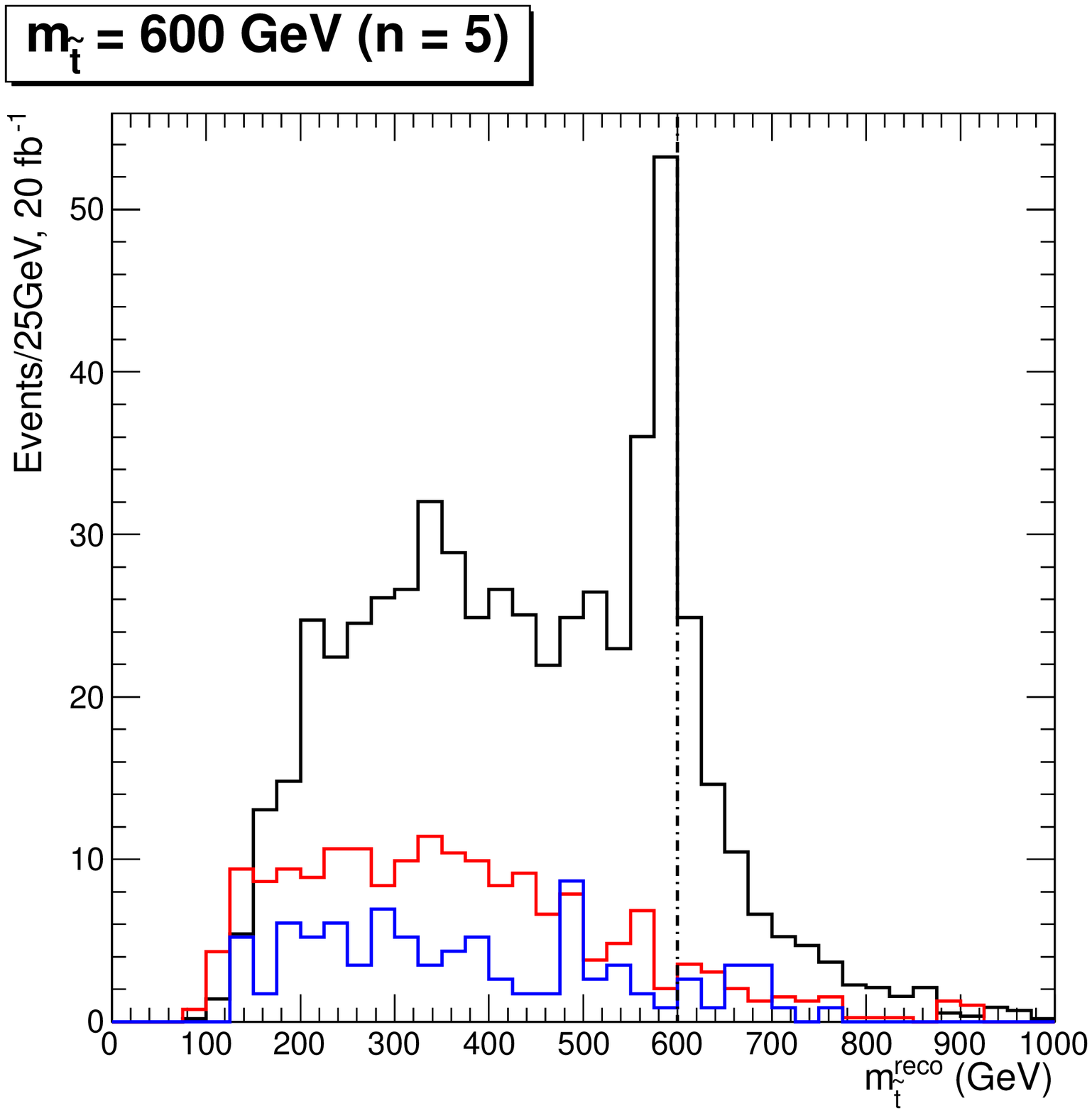}
\epsfxsize=0.27\textwidth\epsfbox{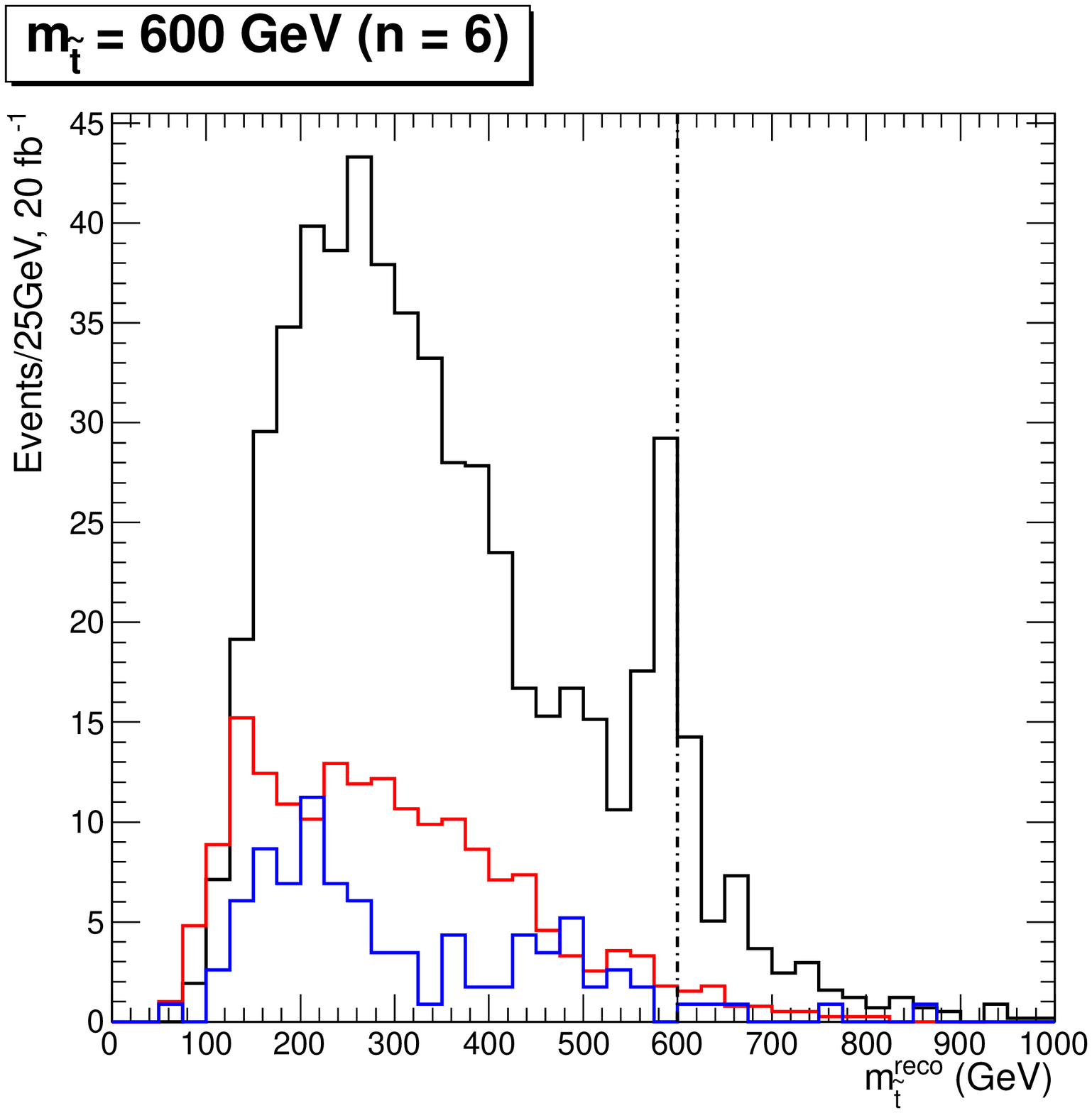}
\\
\epsfxsize=0.27\textwidth\epsfbox{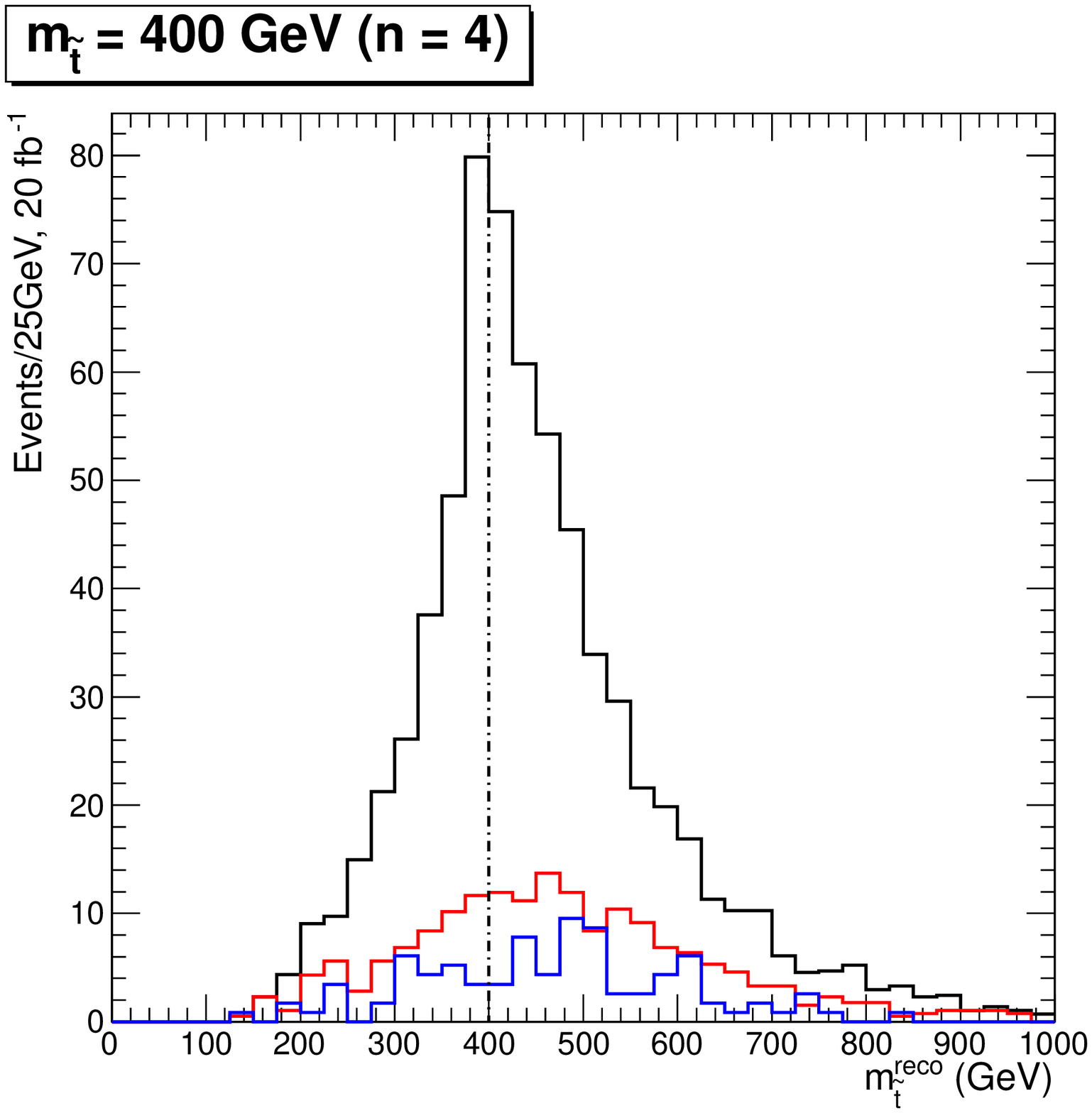}
\epsfxsize=0.27\textwidth\epsfbox{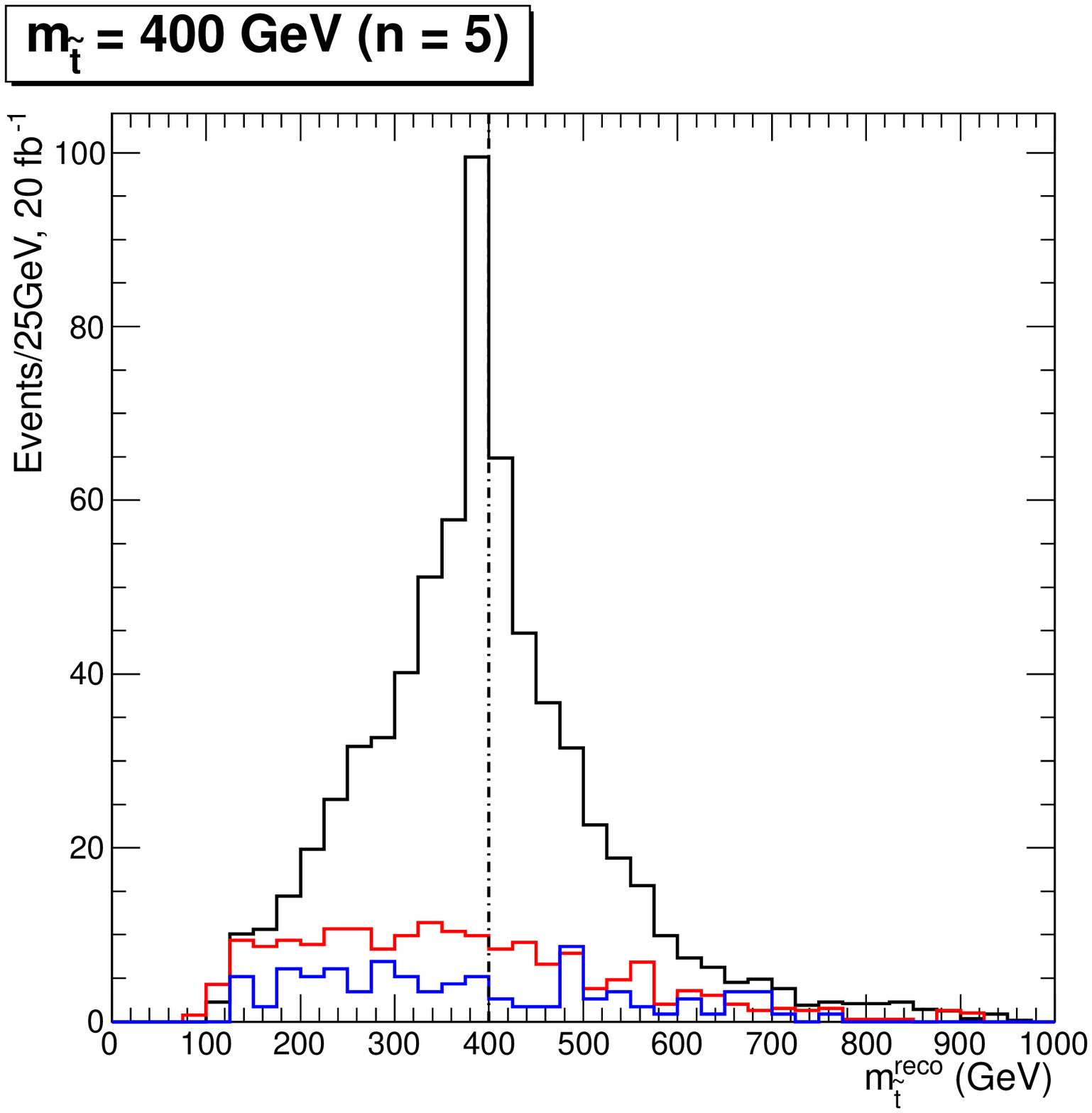}
\epsfxsize=0.27\textwidth\epsfbox{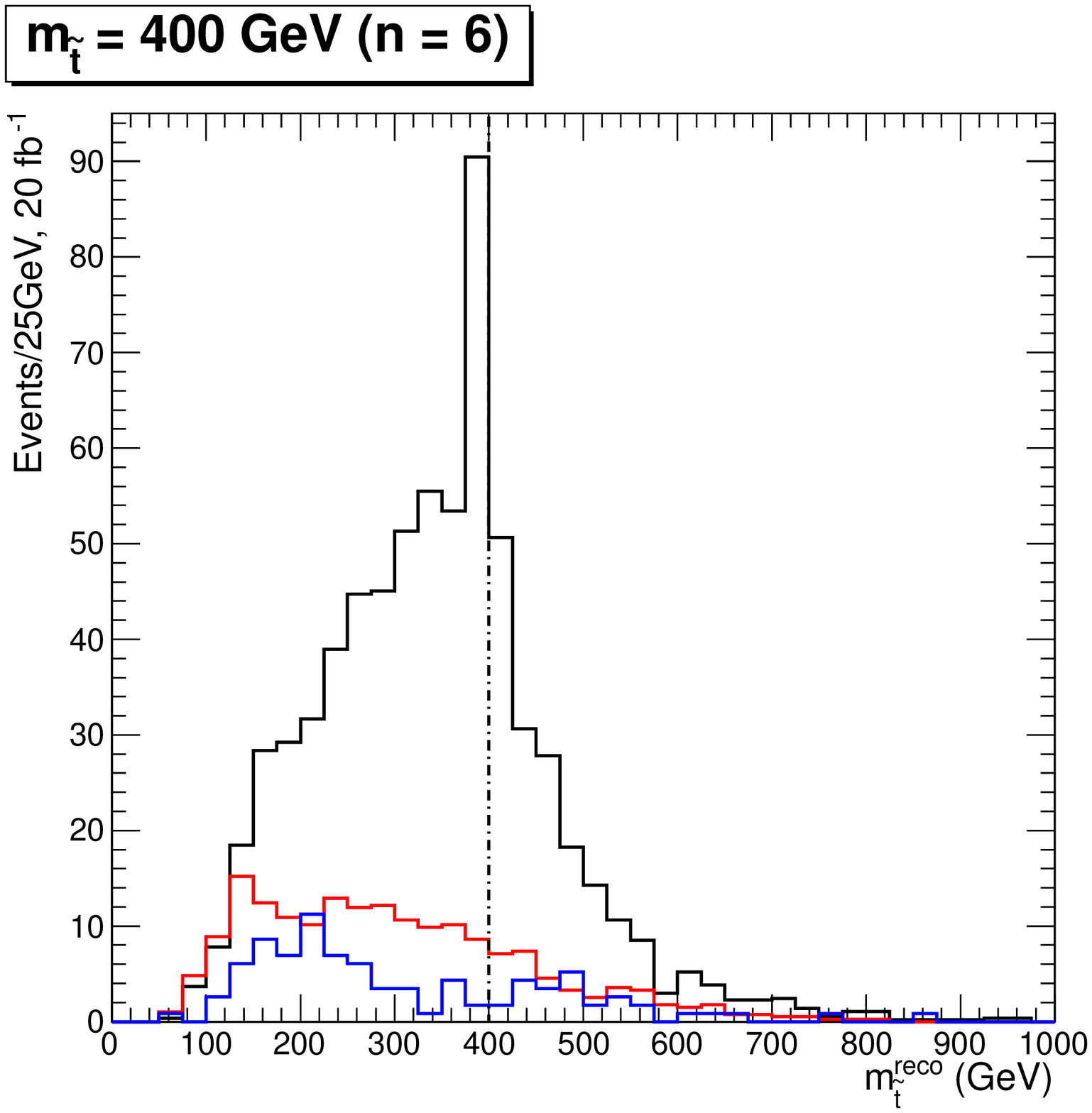}
\\
\epsfxsize=0.27\textwidth\epsfbox{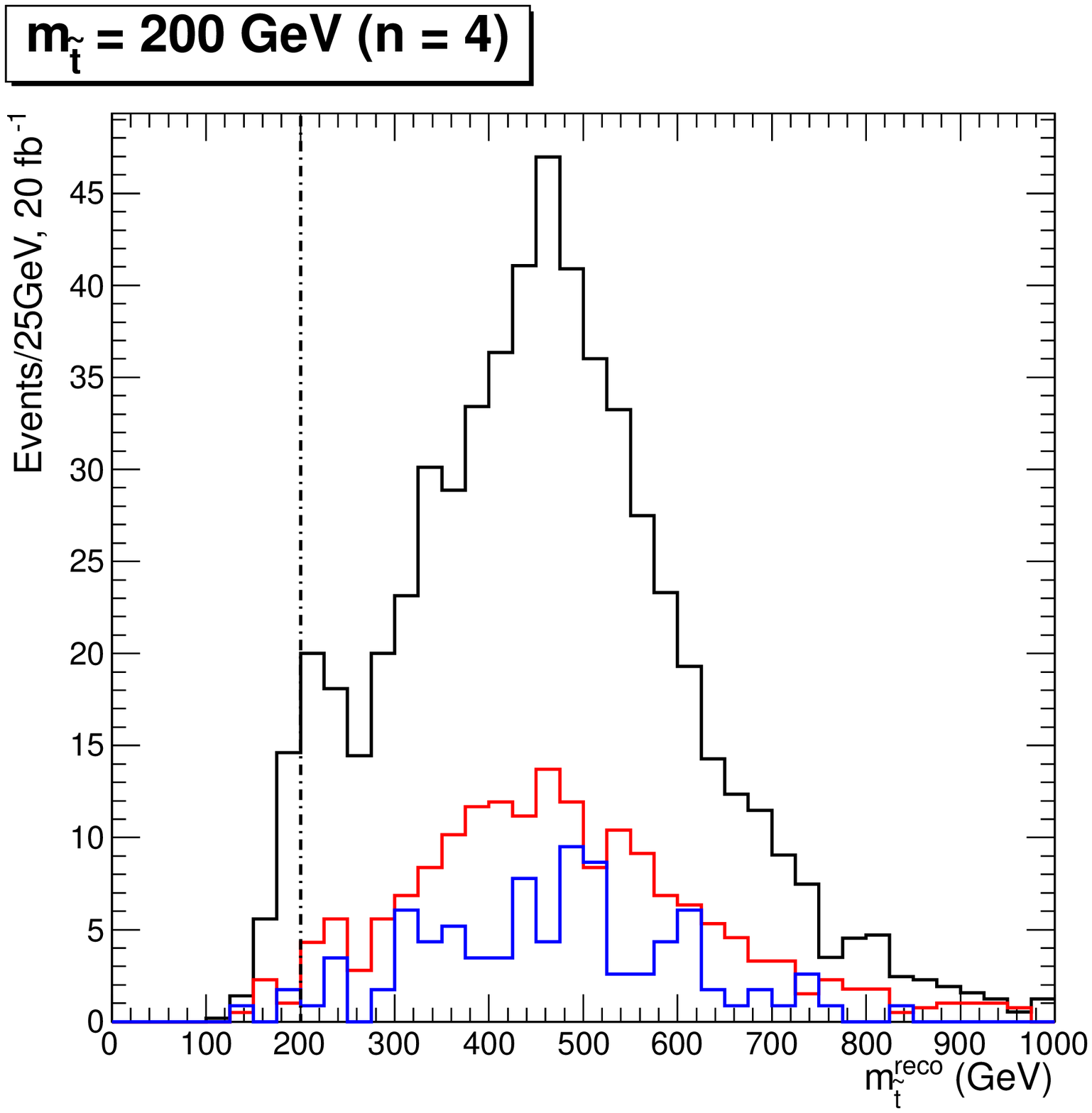}
\epsfxsize=0.27\textwidth\epsfbox{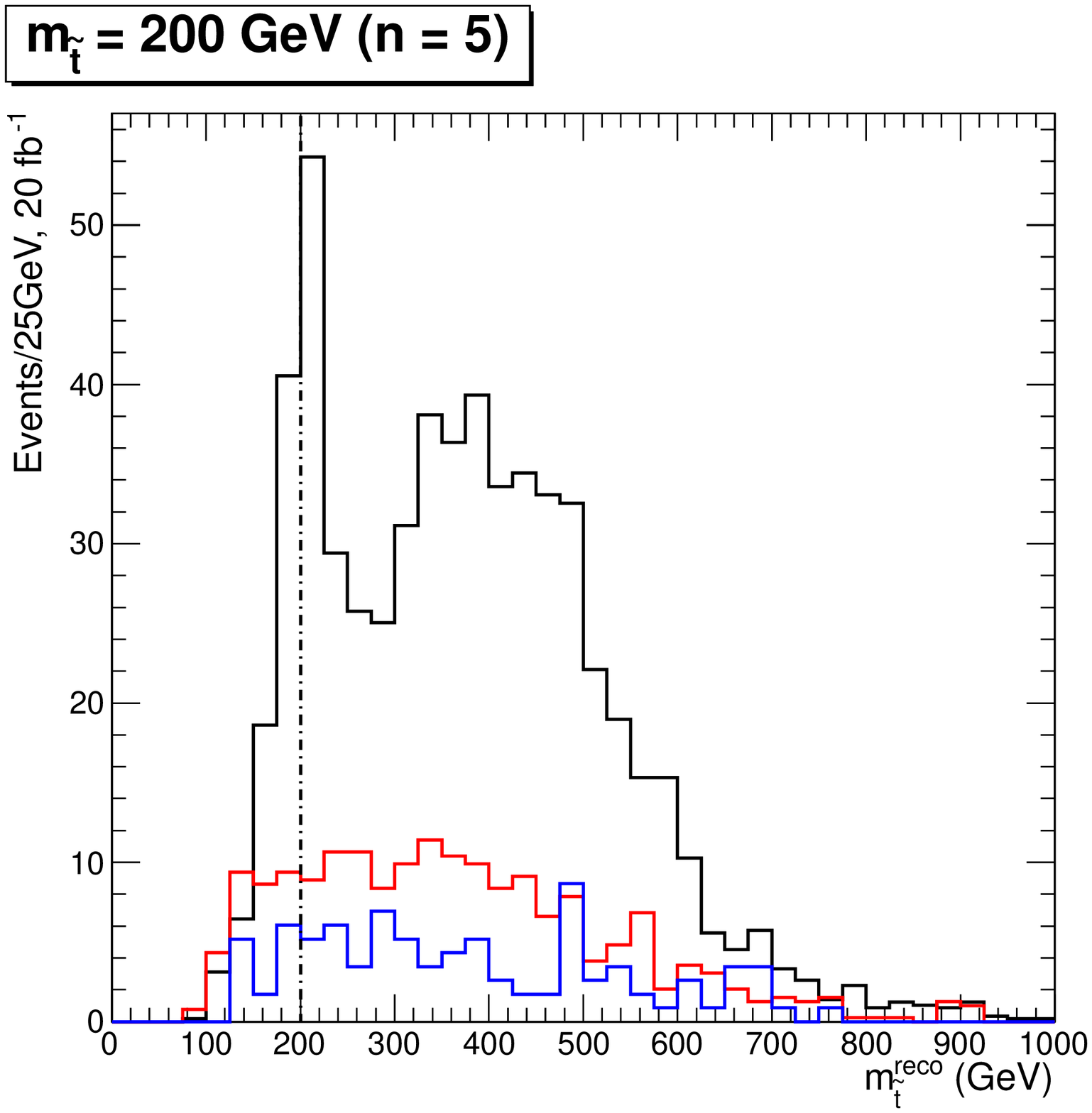}
\epsfxsize=0.27\textwidth\epsfbox{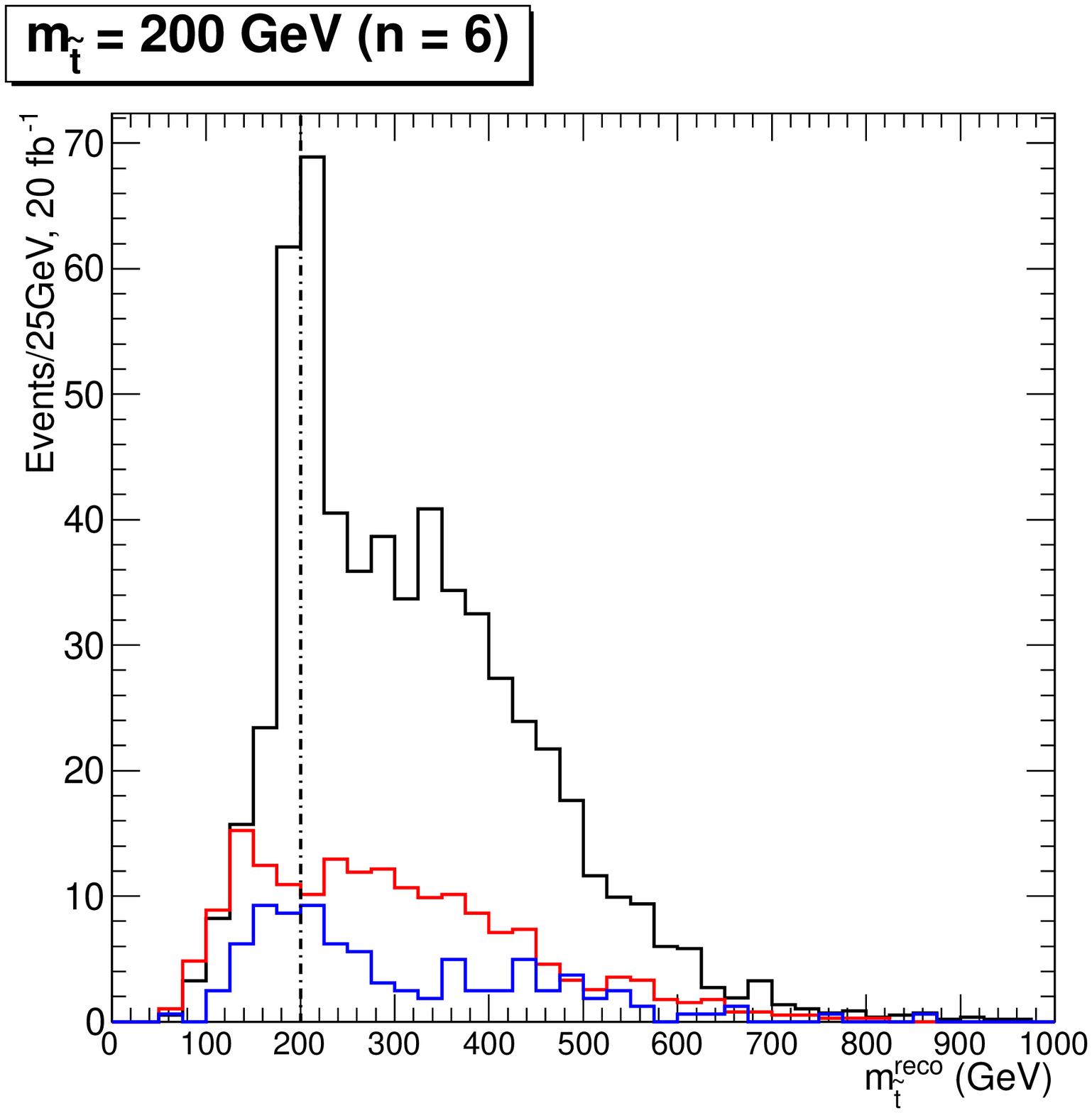}
\caption{Reconstructed stop mass distributions using our traditional jet analysis for 800~GeV Dirac gluinos with 20~fb$^{-1}$ at LHC8.  Black histograms are signal, red histograms are $t\bar t$+jets background, and blue histograms are $W$+jets background.  We consider three different stop masses and three different choices for the number of jets used in the reconstruction.  Rows correspond to 600~GeV stops (top), 400~GeV stops (middle), and 200~GeV stops (bottom).  Columns correspond to reconstructions using the $n$ leading jets with $n =$ 4 (left), 5 (middle), and 6 (right).  We restrict to events with $N_j \ge 7$ and $H_T > 1200$~GeV.  Vertical dashed lines represent the truth stop masses.  (All plots are unstacked.)}
\label{fig:BumpTraditionalJetAnalysis}
\end{center}
\end{figure}

Naively, one would conduct such a search over all available jets.  However, we have found that very frequently, the pairs with the closest mass (either relatively or in GeV) are not the correct ones.  Alternatively, we could restrict ourselves to only the hardest four jets.  This works well in some cases, but not always.  In fact, we have found that the appropriate number of jets to pick depends on the size of the hierarchy between $m_{\stop}$ and $m_{\gluino}$.  We illustrate the situation in Fig.~\ref{fig:BumpTraditionalJetAnalysis}.  Here, we pick from amongst the hardest $n$ leading jets, where $n$ = 4, 5, or 6.  From amongst all partitionings into two pairs, we keep the one where the absolute difference in pair masses is the smallest, and take the average of the two masses.  We see that for larger $m_{\stop}/m_{\gluino}$, we obtain better results with smaller $n$, whereas the reverse trend occurs for smaller $m_{\stop}/m_{\gluino}$.  In order to minimize the amount of spurious reconstruction, we therefore recommend a search that uses a sequence of different $n$'s to cover different ranges of mass ratios.  Note that, even though the choice of $n$ also sculpts the background distributions of the average pair mass in different ways, the signal peak typically remains distinct in shape.

In the following, to estimate our possible reach, we always pick the $n$ that optimizes our signal significance.  To exploit the peak feature's ability to improve $S/B$, we apply a mass window cut of $\pm 20\%$ around the nominal stop mass.
%\beq
%m^{\rm reco}_{\tilde t} \in [m_{\tilde t} - 20\%, \, m_{\tilde t} + 20\%]~.
%\eeq

%------------------------------------------------
\subsection{Bump hunting with jet substructure}
%------------------------------------------------

When the stop is light relative to the gluino, the stop and top quark can both 
be boosted in the decay, and our jet substructure analysis becomes appropriate.  
While it is somewhat rare to find tops or stops fully merged into small jets 
for the gluino masses accessible to LHC8, jet substructure methods very conveniently 
organize the large amount of hadronic information in the event.\footnote{At LHC14, 
jet merging will become an unavoidable issue.  Techniques like the ones we 
discuss here will then become crucial.  Searches in this regime may also 
benefit from a different jet counting criterion, or perhaps even a {\it subjet} 
counting criterion, though this is to some degree implicit in the taggers that we 
employ.  We also comment that the relatively good parton separations at LHC8 
allow us to continue to use traditional lepton isolation.  Dedicated non-isolated 
lepton techniques~\cite{Thaler:2008ju, Rehermann:2010vq} will be required for heavier gluinos.}  
In particular, these methods implicitly exploit the simplifications in combinatorics that occur when stops and tops become boosted, as each will typically beam all of its decay products into localized regions in the detector.  It is also exactly in this regime that our traditional style search becomes least effective.  When the stops are lighter, they take up a smaller fraction of the gluino decay's energy, and jets from the top decays become more difficult to disentangle by simply ranking all jets in $p_T$. 

After applying our baseline $(N_j,H_T)$ cuts described above, we recluster 
the event into $R = 1.5$ ``fat-jets'' using the Cambridge/Aachen 
algorithm~\cite{Dokshitzer:1997in,Wobisch:1998wt}.  Since stop decays are 
two-prong, we can apply the common BDRS substructure 
algorithm~\cite{Butterworth:2008iy}, which first identifies two subjets 
within a fat-jet, and then filters soft and diffuse radiation out of them.  
The mass of the remaining particles is our stop candidate mass.

At this stage, if we simply picked a random fat-jet, we would have some chance of capturing a complete stop decay, a complete hadronic top decay, a part of a decay, or complete or partial decays contaminated by other hard jets.  It is also rare, at least for the mass ranges that we consider, that both stops are completely caught in individual fat-jets, so we cannot reliably apply the nearby-mass trick used in the traditional jet analysis.

To increase our chances of picking a fat-jet that actually corresponds to a fully contained stop, we therefore propose the following.  First, we try to identify a fat-jet corresponding to the hadronic top, characterized by a three-prong system with mass near $m_t$. To do so, we make a first pass over all fat-jets with the HEPTopTagger \cite{Plehn:2009rk, Plehn:2010st}. Amongst all fat-jets passing this tagger (if any), we select the one whose mass is closest to $m_t$\footnote{We have checked that the chance of a stop-jet being mischaracterized as a top-jet is very small (at the percent level) even if $m_{\stop} \simeq m_t$.  The two-body decays of stops lack the additional substructure sought out by top taggers, such as the presence of a subjet doublet consistent with a hadronic $W$ boson.}, and remove it from the list of fat-jets.
Amongst the remaining fat-jets, we take the one with the highest $p_T$ as a stop jet candidate. This is the most likely to consist of a fully-contained stop decay, rather than a fragment of a stop or top decay (either of which only acquires some fraction of the parent energy). 

\begin{figure}[t]
\begin{center}
\epsfxsize=0.44\textwidth\epsfbox{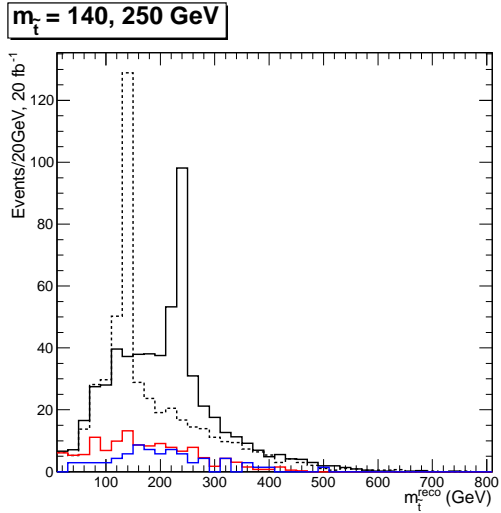}
\caption{Reconstructed stop mass distributions using jet substructure for 800~GeV Dirac gluinos with 20~fb$^{-1}$ at LHC8.  Black histograms are signals with 250~GeV (solid) and 140~GeV (dashed) stops, the red histogram is the $t\bar t$+jets background, and the blue histogram is the $W$+jets background. We restrict to events with $N_j \ge 7$ and $H_T > 1200$~GeV.  (The plot is unstacked.)}
\label{fig:BDRSstopjets}
\end{center}
\end{figure}

We run the BDRS procedure in the default setting on the selected stop jet candidate.  In the rare cases where it fails the procedure, the event is discarded.  The stop mass is then defined as the jet mass after filtering, which is plotted in Fig.~\ref{fig:BDRSstopjets} for some example signals and the SM backgrounds.  While the rate of misreconstruction is not negligible, we nonetheless obtain very narrow stop peaks on top of fairly featureless backgrounds.  As in the traditional analysis, we define a mass window within $\pm20\%$ of the nominal stop mass, and run a refined counting experiment.

%------------------------------------------------
\subsection{Comparison of methods and final results}
%------------------------------------------------

\begin{figure}[t]
\begin{center}
\epsfxsize=0.44\textwidth\epsfbox{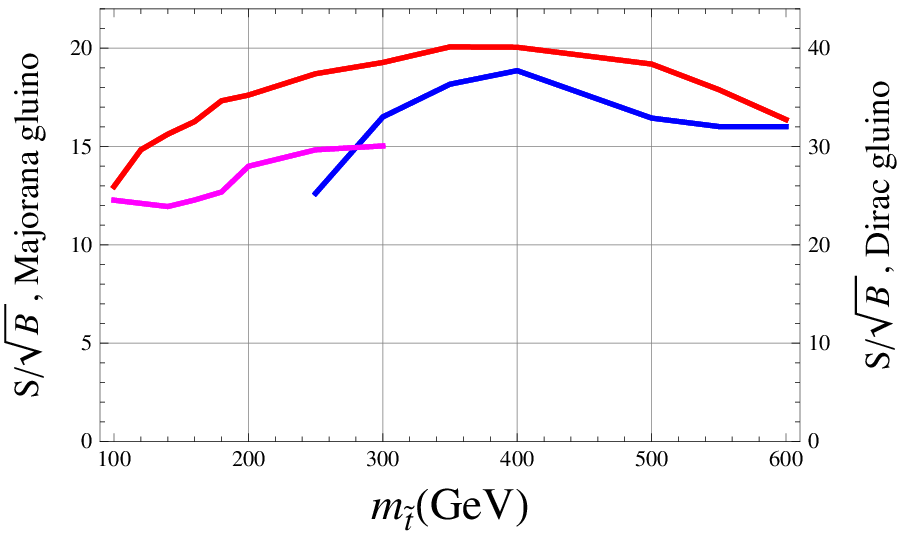}\hspace{5mm}
\epsfxsize=0.44\textwidth\epsfbox{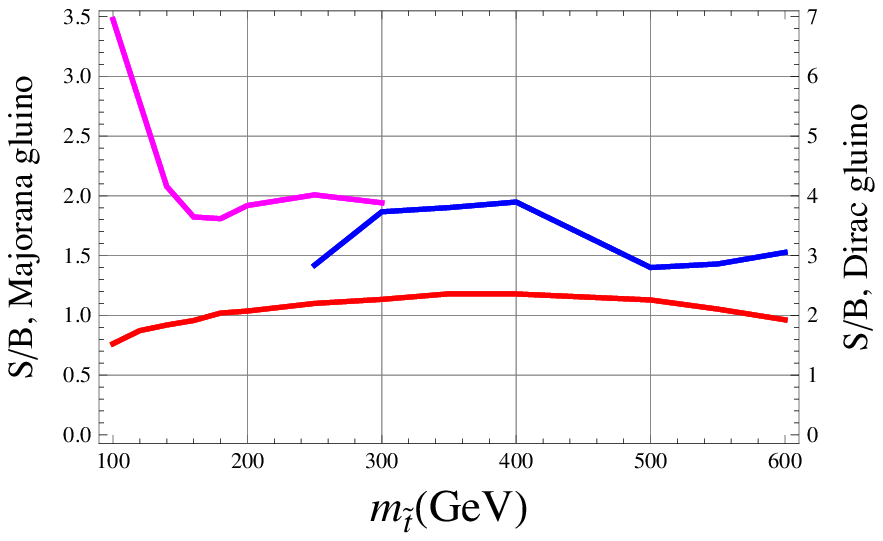}
\caption{The significance ($S/\sqrt{B}$, left) for 20 fb${}^{-1}$ integrated luminosity, and signal/background ratios ($S/B$, right), for an 800~GeV gluino and varying stop masses at LHC8.  The red curves are the simple $(N_j,H_T)$ analysis, the blue curves are with stop peaks reconstructed using traditional jets, and the pink curves use jet substructure.  We have used the cuts given in Table~\ref{tab:800}.}
\label{fig:800}
\end{center}
\end{figure}

With our methods and cuts now determined, we estimate the performance of the different analyses.  We start with a detailed look at the case of an 800~GeV gluino, which we used above to compare several kinematic distributions.  In Fig.~\ref{fig:800}, we see the $S/\sqrt{B}$ and $S/B$ obtainable from all three analyses.  While the basic $(N_j,H_T)$ analysis tends to give somewhat better statistical significance, reconstructing the stops and adding in mass window requirements can improve $S/B$ by as much as a factor of 3.5.  We also see how the substructure based search nicely takes over at low stop masses, where the traditional jet analysis falls off in effectiveness.  The crossover for this gluino mass occurs roughly at $m_{\stop} = 300$~GeV.  We list a complete summary of our final signal and background cross sections for the 800~GeV gluino analysis in table~\ref{tab:800}.  

\begin{figure}[tp]
\begin{center}
   \includegraphics[width=3.2in]{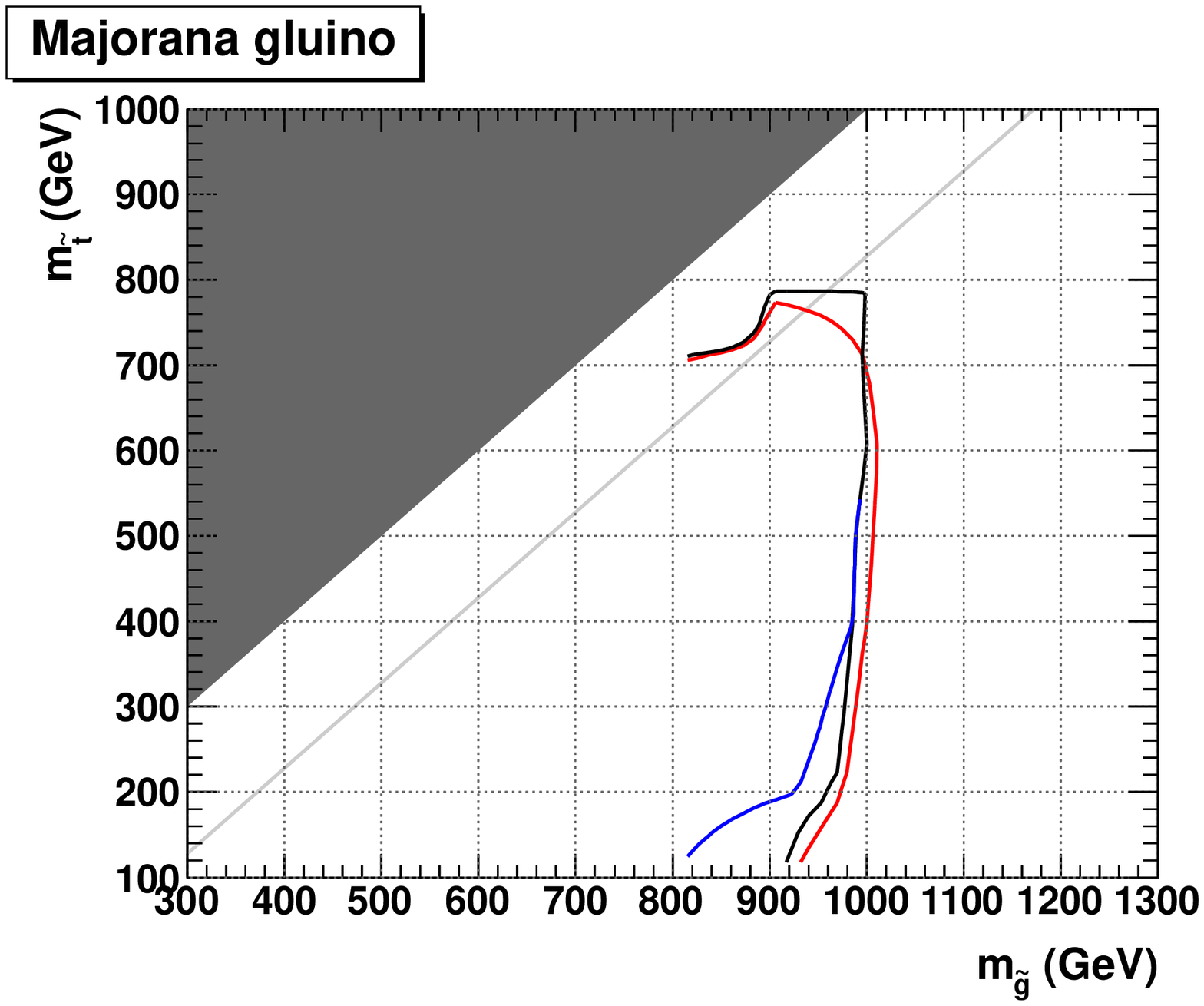}
   \includegraphics[width=3.2in]{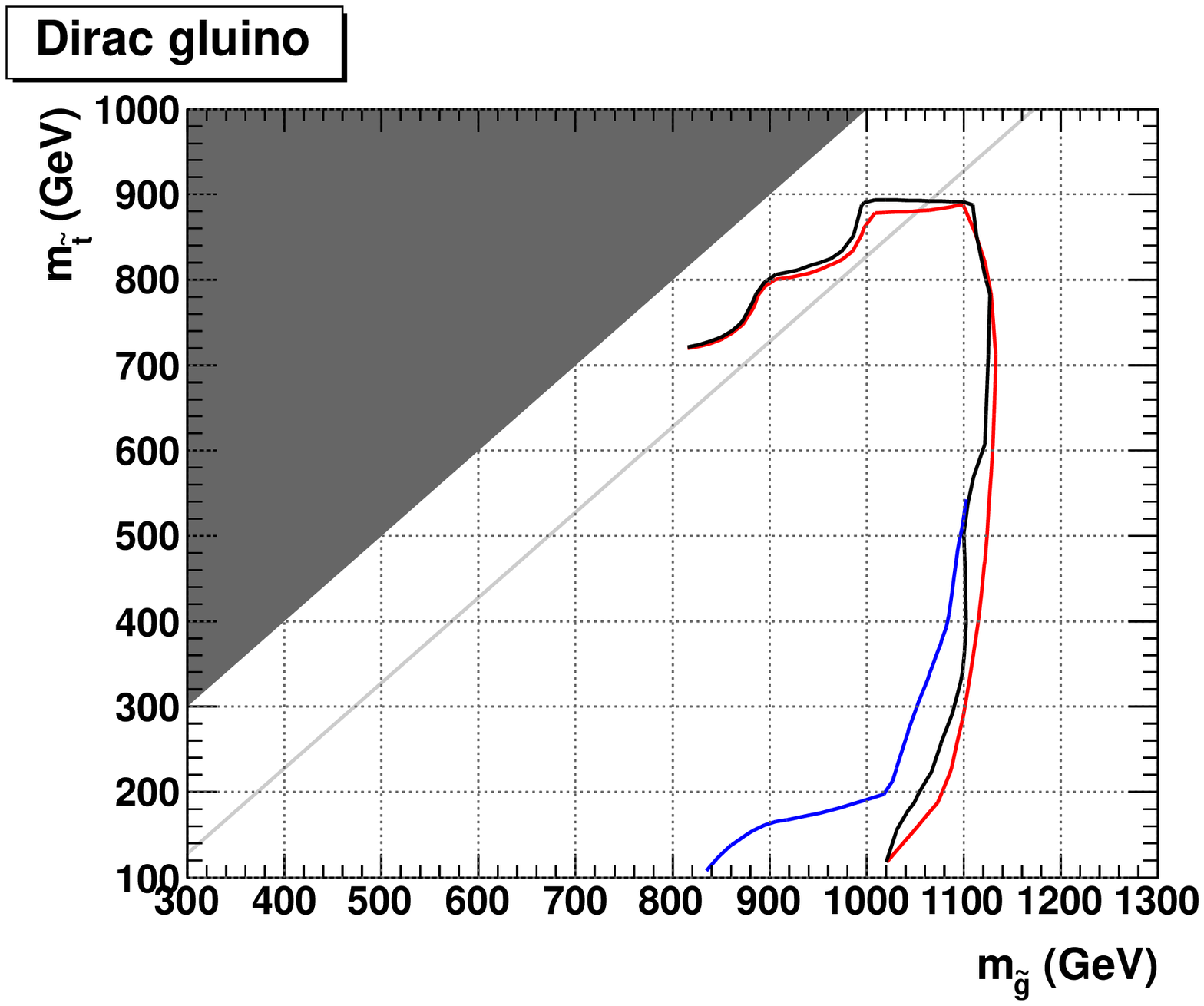}
\caption{Discovery potential ($S/\sqrt{B} \ge 5$) for a Majorana (left) and Dirac (right) gluino with 20~fb$^{-1}$ at LHC8.  The red line is our simple $(N_j,H_T)$ cut-and-count style search.  The black line is the better of our two searches using mass windows around the reconstructed stop peaks (traditional and substructure), which improve discrimination against backgrounds.  The blue line is traditional-only, without substructure.  We do not consider regions with a $\gluino$ LSP, indicated with dark gray.  The light gray line indicates $m_{\gluino} = m_{\stop} + m_t$.}
\label{fig:discovery}
\end{center}
\end{figure}

\begin{table}[tbp]
\centering
{\scriptsize
\begin{tabular}{|c|c|c|c|c|c|c|c|c|}  
\hline
\multicolumn{9}{|c|}{\normalsize $m_{\tilde g} = 800 \ {\rm GeV}$, LHC8 \strut}\\ 
\hline
                   & $m_{\tilde t}$   & 100 GeV & 150 GeV & 200 GeV  & 250 GeV & 300 GeV & 400 GeV & 600 GeV  \\ 
\hline
% Simple $(N_j,H_T)$ & Majorana    & 21.8 fb  & 26.5 fb  & 29.8 fb  & 31.9 fb  & 32.7 fb  & 34.3 fb  & 28.3 fb  \\ \cline{2-9}
 Simple $(N_j,H_T)$ & Majorana       & 11 fb  & 13 fb  & 15 fb  & 16 fb  & 16 fb  & 17 fb  & 14 fb  \\ \cline{2-9}
 $N_j \ge 7$        & Dirac          & 22 fb  & 27 fb  & 30 fb  & 32 fb  & 33 fb  & 34 fb  & 28 fb  \\ \cline{2-9}
 $H_T>1200\gev$     & $t\bar t$+jets & \multicolumn{7}{c|}{9 fb}    \\ \cline{2-9}
                    & $W$+jets       & \multicolumn{7}{c|}{4.5 fb}   \\ 
\hline \hline
                    & $n$ leading jets & \multicolumn{3}{|c|}{-} & $n=6$  & $n=6$  &  $n = 5$  & $n = 4$ \\ \cline{2-9}
%$(N_j,H_T)$ + Traditional $m_{jj}$ reco     & Dirac & \multicolumn{3}{|c|}{-} & 11.2 fb & 15 fb & 18.5 fb & 17.1 fb \\ \cline{2-9}            
 $(N_j,H_T)$ + Traditional $m_{jj}$ reco                     & Majorana        & \multicolumn{3}{|c|}{-}  & 5.6 fb &  7.5 fb & 9.3 fb & 8.6 fb \\ \cline{2-9}            
 $0.8 m_{\tilde t} < m^{\rm reco}_{\tilde t} < 1.2 m_{\tilde t}$  & Dirac           & \multicolumn{3}{|c|}{-}  &  11 fb &   15 fb &  19 fb &  17 fb \\ \cline{2-9}            
                                                            & $t\bar t$+jets  & \multicolumn{3}{|c|}{-}  & 2.4 fb &  2.8 fb & 3.0 fb & 3.3 fb \\ \cline{2-9}
                                                            & $W$+jets        & \multicolumn{3}{|c|}{-}  & 1.4 fb &  0.9 fb & 1.2 fb & 1.8 fb \\ 
\hline \hline
%$(N_j,H_T)$ + Jet substructure & Dirac & 6.4 fb  & 10.1 fb  & 11.9 fb  & 11.6 fb  & 10.1 fb  &  \multicolumn{2}{|c|}{-}  \\ \cline{2-9}
                                                         & Majorana      & 3.2 fb &  5.6 fb &  6.0 fb &  5.8 fb &  5.0 fb &  \multicolumn{2}{|c|}{-}  \\ \cline{2-9}
$(N_j,H_T)$ + Jet substructure                           & Dirac         & 6.4 fb & 10 fb & 12 fb & 12 fb & 10 fb &  \multicolumn{2}{|c|}{-}  \\ \cline{2-9}
$0.8m_{\tilde t} < m^{\rm reco}_{\tilde t} < 1.2m_{\tilde t}$ & $t\bar t$+jets & 0.9 fb & 1.5 fb & 1.6 fb & 1.6 fb  & 1.3 fb &    \multicolumn{2}{|c|}{-}  \\ \cline{2-9}
                                                        & $W$+jets       & 0.1 fb & 0.7 fb & 1.3 fb & 0.8 fb  & 0.6 fb &    \multicolumn{2}{|c|}{-}  \\ 
\hline
\end{tabular}
}
\caption{LHC8 cross sections for an 800 GeV gluino and for backgrounds, after the different analysis cuts.  The total cross section for the Majorana (Dirac) signal is 43~fb (86~fb) in the $l$+jets channel before cuts.  We assume $BR(\gluino \rightarrow t \stop)=1$.}
\label{tab:800}
\end{table}

In Fig.~\ref{fig:discovery}, we show the boundary in the $(m_{\gluino},m_{\stop})$ plane of models that would be discoverable by the end of 2012.  Masses up to 1~TeV are attainable for Majorana gluinos, and beyond 1.1~TeV for Dirac gluinos.  The results are not very sensitive to the stop mass, except as we approach the line $m_{\gluino} = m_{\stop}+m_t$.  In particular, we can again see the substructure-based search taking over at lower stop masses.  The equivalent exclusion contours (not shown), would move up to roughly 1.1~TeV and 1.2~TeV, respectively.

In all of our analyses, we have assumed a 100\% branching ratio for the gluino decaying to a top-stop pair. If other decay channels are open, more dedicated reconstruction methods would help to maximize the discovery potential.  Nevertheless, these decay channels can share similar features with the top-stop channel, and many of the strategies discussed above still apply with little or no modification.  For example, if a sbottom is present below the gluino mass and above the stop mass, we obtain the same final state particles through the decay chain $\gluino\rightarrow b\sbottom \to b(W\stop)$. As an example,  we have tested the mass point $(\mgluino, \msbottom, \mstop)=(800, 300, 200)$~GeV and considered the case $\gluino\gluino\rightarrow t\stop b\sbottom$. We let the $W$ from the top decay hadronically and the one from the sbottom decay leptonically, and apply the same ($N_j, H_T$) cuts as in Table \ref{tab:800}. This results in only a slightly different signal efficiency of 31\%, from 35\% for the case when both gluinos decay to a top and a stop.

\section{Conclusions and Outlook}
\label{sec:conclusions}

%%%%%%%%%%%%%%%%%%%%%%
%  Conclusions and Outlook
%%%%%%%%%%%%%%%%%%%%%%

From the perspective of natural supersymmetry, it may not be surprising that 
superpartners have so far been elusive at the LHC.  On the one hand, this 
framework places priority on light third generation squarks to maintain naturalness, 
while the other squarks and sleptons could simply be too heavy to produce.  
On the other hand, a fully general approach demands 
additional symmetries beyond R-parity to maintain proton stability, and makes 
R-parity violation a perfectly viable option.  We should then take seriously the 
possibility that stops are light, and can decay directly to two jets via the 
$u^cd^cd^c$ superpotential operator.  In this paper, we have approached the 
difficult task of finding $\stop \to jj$ in the context of gluino decays.  
We focused on the power of $l$+jets searches with large jet multiplicity, and 
established the possibility of reconstructing the stop mass peaks despite the 
combinatoric challenges.  In addition to the advantageous kinematic handles and 
the chance to discover two particles simultaneously, searching for stops in gluino 
production also provides a crucial opportunity to infer the stop's identity because 
it would be produced in association with top quarks. 

Current searches for new physics are not optimized for this process.  Nevertheless, 
non-trivial bounds can be inferred, and we have estimated the current allowed 
parameter space based on these existing searches.  If we take the default 
assumption that the gluino is Majorana, then SS dilepton searches apply, and 
limits up to almost 800~GeV can be placed for stops lighter than about 350~GeV.  
If instead the gluino is Dirac, which would alleviate many low-energy constraints 
on RPV while potentially depleting the SS dilepton signal, then the most powerful 
limits actually come from the ATLAS black hole search.  The limit exceeds 600~GeV
for $m_{\stop} \simeq 100$~GeV, and becomes weaker for heavier stops.

In either case, we have found that even a simple cut-and-count based search for 
high-$N_j$, high-$H_T$ $l$+jets events would extend these limits. One could already 
exclude Majorana (Dirac) gluinos as heavy as 1.0~TeV (1.1~TeV) with this method, 
using an integrated luminosity of 5~fb$^{-1}$ at LHC8.  With 20~fb$^{-1}$, these 
masses are at the discovery boundary, and exclusions move out to 1.1~TeV (1.2~TeV).

To reconstruct the stop in gluino pair events, we proposed two approaches.  
The first uses traditional jets, and attempts to reconstruct both stops by 
studying pairs of pairings amongst the hardest 4, 5, or 6 jets in each event.  
This approach works well when $m_{\stop}/m_{\gluino} \gtrsim 1/3$.  Our second 
approach uses jet substructure.  We identify fat-jets consistent with a two-body 
decay, and also reject fat-jets that look more consistent with three-body top decays.  
This approach works well when $m_{\stop}/m_{\gluino} \lesssim 1/3$.  We demonstrated 
that the two approaches facilitate reliable stop mass peak reconstruction in 
their respective regimes of validity, and that there is a healthy overlap 
between these regimes.  The additional step of reconstructing the stop peak is not 
only useful for characterizing an excess in high-$N_j$, high-$H_T$ events, 
but provides us with an important shape discriminant against the backgrounds 
and improves $S/B$.  We expect that it will have an important role to play in 
controlling systematic uncertainties in these searches.

Assuming that such a dijet resonance bump is discovered, we can also immediately 
begin to understand whether it is indeed a top partner.   The main goal would 
be to establish the presence of tops quarks in the signal.  Such an endeavor 
would have to be undertaken with great care, since the main background is 
itself $t\bar t$+jets.  However, we have seen that it is possible to achieve 
substantial $S/B$, making the model discrimination much easier.  Of course, 
the presence of an accompanying leptonic $W$ boson is easy to infer based on the 
transverse mass of the lepton and $\met$.  This could be used to rule out the 
presence of a new invisible particle produced in association with the dijet 
resonances, for example by showing that the resonance disappears for 
$m_T(l,\met) > m_W$.  It is somewhat more difficult to reconstruct the top quarks 
themselves, owing to the busy jet environment.  Probably the best strategy is to 
try to find the leptonic top peak, reducing the combinatorics by demanding $b$-tags 
from the jets (or subjets) that were not used in stop reconstruction.  Since the 
tops are typically at least somewhat boosted, combining the leptonic $W$ with 
the closest $b$-tagged jet would likely be promising.  A more global kinematic 
$\chi^2$ style reconstruction, which attempts to reassemble every intermediate 
particle in the event, would also be interesting to explore.

Another useful way to further characterize the signal would be to fully reconstruct 
the gluino mass peaks.  We already do this in an approximate way with $H_T$, and 
a slightly more careful analysis that uses $H_T$ shape information would 
unambiguously tell us the gluino mass.  However, it should also be possible to 
reconstruct one gluino at a time if we are able to also reconstruct top quarks 
in the event.

While we have found very similar kinematics for Majorana and Dirac gluinos,
we point out that these may already be distinguishable based on total rate, though such a
comparison would depend on $BR(\gluino \to t\stop)$.  A more definitive test would
be a measurement of the relative rates of same-sign and opposite sign dilepton signals, which becomes
feasible with more data.

Naturalness suggests that gluinos are not much heavier than a TeV if they 
are Majorana, and can be just factor of $\cO(2)$ heavier if they are Dirac.  
While our estimated LHC8 coverage 
only just reaches the TeV-scale, the remaining parameter space will likely be 
covered at the higher energy run.  The search strategies that we proposed in 
this paper should straightforwardly extend beyond a TeV, in particular the search 
based on jet substructure.

\appendix

\section{Simulation and Reconstruction Details}
\label{app:simulations}

%%%%%%%%%%%%%%%%%%%%%%
%  Appendix - simulations details
%%%%%%%%%%%%%%%%%%%%%%

\label{app:simulation}
We generate all of our signal and background samples with 
{\tt MadGraph5} {\tt v1.4.7}~\cite{Alwall:2011uj}.  The gluino pair signal 
samples are based on our own simplified UFO~\cite{Degrande:2011ua} models, which are in 
turn derived from the publicly available {\tt RPVMSSM} model~\cite{Fuks:2012im}.
Gluinos can be either Majorana or Dirac, and we assume pure-RH stops.  
We normalize LHC7 signal cross sections to their NLO+NLL values~\cite{Beenakker:2011fu}, 
and assume $BR(\gluino \to t \stop^* \; {\rm or} \; \bar t \stop) = 1$.  We apply a rescaling factor of 
2 for Dirac gluinos, and extrapolate all cross sections to LHC8 by using 
leading-order ratios between LHC8 and LHC7.

Our background samples, generated only for LHC8, consist of $t\bar t$+jets 
matched up to two jets and $W$+jets matched up to four jets.  We use cone-MLM 
matching~\cite{Alwall:2007fs} with $R = 0.4$ and $p_T = 30$~GeV.  For both backgrounds we 
apply an approximate $K$-factor of 2.  While we would ideally 
match $t\bar t$+jets up to four jets and $W$+jets up to seven or eight jets, these are too 
computationally intensive.  However, we do compare our nominal samples to 
simulations requiring fewer numbers of matched jets, and obtain similar results

Backgrounds are automatically processed through {\tt PYTHIA6}~\cite{pythiamanual}.  
The nonstandard color structure of the RPV vertex requires us to instead use 
{\tt Pythia8}~\cite{Sjostrand:2007gs} for the signal, in order to obtain sensible 
showering and hadronization.\footnote{It is highly likely that the stops in fact hadronize before they decay, which effectively shields this color structure.  However, we do not expect that these details have any impact on our results.  We note that it is also possible to force the stop to decay to dijets within {\tt Pythia8} (and in principle in {\tt PYTHIA6}), though we have not utilized this option.}  We do not further process the hadron-level event using a detector model.

Our basic event reconstruction proceeds as follows.  We first demand exactly 
one isolated lepton with $p_T(l) >$ 20 GeV and $|\eta(l)| < 2.5$.  A lepton is
considered isolated if the surrounding hadronic activity within 
a cone of size $R = 0.3$ satisfies $p_T(l)/(p_T(l)+p_T({\rm cone})) > 0.85$. 
We then cluster the rest of the particles in the event into $R$ = 0.5 jets with 
the anti-$k_T$ algorithm~\cite{Cacciari:2008gp} using {\tt FastJet}~\cite{Cacciari:2005hq}, 
accepting only jets with $p_T(j) >$ 40 GeV and $|\eta(j)| <$ 2.5.  At this stage, we
simply count jets to define $N_j$ and define the variable $H_T$ as in equation (\ref{eq:HT}).

%%%%%%%%%%%%%%%%%%%%%%%%%%%%%%%%%%%%%%%%%%%%%%%%%%%%%%%%%%%%%%%%%%%%%%%%

\acknowledgments{We thank Keith Rehermann for collaboration in the early phase of this work.  We thank Johan Alwall 
and Olivier Mattelaer for help in simulating RPV process with {\tt MadGraph5}.  We are grateful to Roberto Contino, Tobias Golling, Yuval Grossman and Liantao Wang for useful discussions.  We also thank Tobias Golling for comments on the draft, and Martin Schmaltz for 
helping to supplement our computer resources.   ZH and MS are grateful to the organizers of the workshop ``BOOST2012'' at Valencia.
ZH was supported in part by DoE grant No.\ DE-FG-02-96ER40969. AK was supported by NSF grant No.\ PHY-0855591.  
MS was supported in part by DoE grant No.\ DE-FG-02-92ER40704, and by the ERC Advanced Grant No. 267985, 
``Electroweak Symmetry Breaking, Flavour and Dark Matter: One Solution for Three Mysteries'' (DaMeSyFla). 
BT was supported by DoE grant No.\ DE-FG-02-91ER40676 and by NSF grant No.\ PHY-0969510 (LHC Theory Initiative).  
AK and BT thank the Aspen Center of Physics, supported by NSF grant No.\ PHY-1066293, where part of this work was done.  }

%%%%%%%%%%%%%%%%%%%%%%%%%%%%%%%%%%%%%%%%%%%%%%%%%%%%%%%%%%%%%%%%%%%%%%%

%%%%%%%%%%%%%%
% References
%%%%%%%%%%%%%%

\bibliography{lit}
\bibliographystyle{apsper}

\end{document}